\documentclass[final,authoryear,3p,times]{elsarticle}
\usepackage[latin9]{inputenc}
\usepackage{amsmath}
\usepackage{amssymb}
\usepackage{graphicx}
\usepackage{esint}

\makeatletter

\providecommand{\tabularnewline}{\\}

 \usepackage{lineno}

\journal{J. of Atmospheric and Solar-Terrestrial Physics}

\begin{document}
\begin{frontmatter}

\title{Does the Sun work as a nuclear fusion amplifier of planetary tidal forcing? \\
A proposal for a physical mechanism based on the mass-luminosity relation}

\author{Nicola Scafetta $^{1}$}

\address{$^{1}$ACRIM (Active Cavity Radiometer Solar Irradiance Monitor Lab)
\& Duke University, Durham, NC 27708, USA.}
\begin{abstract}
Numerous empirical evidences suggest that planetary tides may influence
solar activity. In particular, it has been shown that: 1) the well-known
11-year Schwabe sunspot number cycle is constrained between the spring
tidal period of Jupiter and Saturn, $\sim$9.93 year, and the tidal
orbital period of Jupiter, $\sim$11.86 year, and a model based on
these cycles can reconstruct solar dynamics at multiple time scales
\citep{Scafetta200}; 2) a measure of the alignment of Venus, Earth
and Jupiter reveals quasi 11.07-year cycles that are well correlated
to the 11-year Schwabe solar cycles; 3) there exists a 11.08 year
cyclical recurrence in the solar jerk-shock vector, which is induced
mostly by Mercury and Venus. However, Newtonian classical physics fails to explain the phenomenon.
Only by means of a significant nuclear fusion
amplification of the tidal gravitational potential energy released
in the Sun, may planetary tides produce irradiance output oscillations
with a sufficient magnitude to influence solar dynamo processes. Here
we explain how a first order magnification factor can be roughly calculated  using an adaptation of the well-known \emph{mass-luminosity relation}
for main-sequence stars similar to the Sun. This strategy yields a
conversion factor between the solar luminosity  and the potential
gravitational power associated to the mass lost by nuclear fusion:
the average estimated amplification factor is $A\approx4.25\cdot10^{6}$.
We use this magnification factor to evaluate the theoretical luminosity
oscillations that planetary tides may potentially stimulate inside
the solar core by making its nuclear fusion rate oscillate. By converting
the power related to this energy into solar irradiance units at 1
AU we find that the tidal oscillations may be able to theoretically
induce an oscillating luminosity increase  from 0.05-0.65
$W/m^{2}$ to 0.25-1.63 $W/m^{2}$, which is a range compatible with
the ACRIM satellite observed total solar irradiance fluctuations.
In conclusion, the Sun, by means of its nuclear active core, may be
working as a great amplifier  of the small
 planetary tidal energy dissipated in it. The amplified signal should be sufficiently energetic
to synchronize solar dynamics with the planetary frequencies and activate
internal resonance mechanisms, which then generate and interfere with
the solar dynamo cycle to shape solar dynamics, as further explained
in \citet{Scafetta200}. A section is devoted to explain how the traditional
objections to the planetary theory of solar variation can be rebutted.
\end{abstract}


\textbf{Please, cite this article as:} Scafetta, N., (2012). Does the Sun work as a nuclear fusion amplifier of planetary tidal forcing? A proposal for a physical mechanism based on the mass-luminosity relation. Journal of Atmospheric and Solar-Terrestrial Physics 81-82, 27-40. http://dx.doi.org/10.1016/j.jastp.2012.04.002 \\

\end{frontmatter}


\section{Introduction}

It is currently believed that solar activity is driven by internal
solar dynamics alone. In particular, the observed quasi-periodic 11-year
sunspot and total solar irradiance (TSI) cycles are believed to be
the result of solar differential rotation, which is modeled in hydromagnetic
solar dynamo models \citep{Tobias}. However, the severe incompleteness
of the current solar theories assuming that the Sun acts as an
\emph{isolated system} is indirectly demonstrated by their inability to
reconstruct the solar variability occurring at multiple time scales, , as also critics of the planetary theory acknowledge \citep{Jager}.
On the contrary, since the 19$^{th}$ century \citep{Wolf1859} a
theory has been proposed claiming that solar dynamics is partially
driven by planetary tides. This theory has never been definitely disproved,
although a plausible physical mechanism has not been discovered yet.

Indeed, a planetary theory of solar variation has  been criticized
for various reasons \citep{Charbonneau}. There are three classical
objections. A first
objection claims that planetary tides alone would not explain solar
variability \citep{Smythe}. A second objection based on classical physics claims that the planetary
tidal forces are too small to modulate solar activity: for example,
tidal accelerations at the tachocline level are about 1000 times smaller
than the accelerations of the convective motions \citep{Jager,Callebaut}. A
third objection claims that traditional concepts in the theory of
stellar structure such the Kelvin-Helmholtz time scale \citep{Mitalas,Stix}
would predict that, because the erratic propagation of the light from
the core to the convective zone requires $10^{4}$ to $10^{8}$ years,
changes in the luminosity production would be smoothed out before
reaching the convective zone, and the smoothed anomaly signal would
not be observable in the final luminosity output. The above three
major objections have prevented solar scientists from further investigating
the issue taking advantage of the larger and more accurate satellite
data about solar activity collected since 1980s, of computer modeling and of theoretical thinking based on modern physics.

On the contrary, the theory that solar activity may be linked to
planetary motion would be supported by a large number of empirical
evidences \citep{Wolf1859,Schuster,Takahashi,Bigg,Jose,Wood65,Wood,Dingle,Okal,Fairbridge,Charvatova90,Charvatova20,Charvatova09,
Landscheidt88,Landscheidt99,Juckett,Hung,Wilson,Scafetta,Perryman,scafett2011}. Recently, \citet{Scafetta200} has shown that it is possible to
reconstruct solar variability with a very good accuracy at the decadal,
secular and millennial scale throughout the Holocene using a model
based on the tidal cycles of Jupiter and Saturn plus the dynamo cycle.
Scafetta's results have strongly rebutted the empirical criticism
by \citet{Smythe}, for example, and have reopened the issue.

The geometrical patterns of the motion of the Sun relative to the
barycenter of the solar system have been used by some of the above
authors to support a planetary influence on solar activity. The complex
wobbling of a star around the barycenter of its solar system is a
well-known phenomenon of stellar motion \citep{Perryman}. Indeed,
\citet{Wolff} have recently proposed that the rotation of the Sun
around the barycenter of the solar system could induce small mass
exchanges that release potential energy. The mass exchange would also
carry fresh fuel to deeper levels and increase solar activity. This
phenomenon would cause stars like the Sun with an appropriate planetary
system to burn somewhat more brightly and have shorter lifetimes than
identical stars without planets. However, the solar barycentric motion
should be understood just as an approximate geometrical proxy of the
forces acting on the Sun. Tidal forces, torques and jerk shocks act
on and inside the Sun, which is not just a point-size body in free
fall. Only these forces can potentially influence solar activity according
to the laws of mechanics, although additional more complex mechanisms cannot be excluded.

Herein, we observe that the continuous tidal \emph{massaging} of the
Sun should be heating the solar core too. Tidal heating, where orbital
and rotational energy is dissipated as heat through internal friction
processes, is a well-known planetary phenomenon \citep{Goldreich,Jackson}.
Particularly, tidal heating effects are macroscopic in the case of
Jupiter's moon Io \citep{Bennett}. In the case of a star, we hypothesize
that tidal heating  modulates the nuclear fusion rate and,
therefore, the total solar irradiance (TSI) output of the Sun. Thus, the planetary
gravitational tidal energy dissipated in the core should be greatly
amplified by the internal nuclear fusion rate response to it. We propose
a rough estimate of both the value of the planetary gravitational
tidal heating of the solar core and of the internal nuclear feedback
amplification factor that would amplify the released gravitational
tidal potential energy into TSI output. Finally, we compare the magnitude
of the planetary tidal induced TSI output variation against the observed
TSI variation. The theoretical results of this paper, which are based on modern physics, would rebut the
second major objection against the planetary-solar theory, which uses arguments based on classical physics alone to claim
that  planetary tides are too small to influence the Sun \citep{Jager}. Indeed, the failure of the 19th century Kelvin-Helmholtz timescale theory claiming that the Sun is about 10 million years old instead of the currently accepted age of 4.7 billion years \citep{Carroll} demonstrates that  classical physics alone does not explain how the Sun works by a large factor. Indeed, the well-known fact that stars are not classical physical systems can invalidate any argument that uses classical physics alone to disprove a planetary tidal influence on the Sun.

The third objection is not explicitly addressed herein. However, the
interior of the Sun is not in a pure hydrostatic equilibrium, but
it is likely crossed by buoyancy gravitation-waves known as g-mode waves
\citep{Garcia}. \citet{Wolff} proposed that g-mode waves may activate
extremely fast upward transport mechanisms of luminosity variation
from the core to the surface. Wave propagation mechanisms would solve
the theoretical problems related to the theorized extremely slow luminosity
diffusion movement occurring in the radiative zone \citep{Mitalas,Stix}.

In Section 2 we summarize some empirical findings suggesting that
planetary tides may influence solar activity. In Section 3 we develop
a simple energetic model argument to demonstrate that such a theory
may be physically plausible. We conclude that the empirical and theoretical
evidences in favor of a planetary influence on solar activity are
too strong to be ignored: in the future there is the need to study
them extensively and include these mechanisms in future solar models.
This would be greatly beneficial to solar physics and climate science
as well \citep{Scafetta200}.

\section{Empirical evidences for a planetary forcing on the Sun}

The possibility that solar cycles are partially modulated by planetary
tidal cycles has been frequently suggested since the 19th century \citep{Wolf1859,Brown,Bendandi,Schuster}.
For example, in a short letter to Carrington, Wolf (1859) proposed
that the variations of sunspot-frequency depend on the influences
of Venus, Earth, Jupiter and Saturn. Brown (1900) noted that the 11-year
sunspot cycle and its multidecadal variation could be linked to the
combined influence of Jupiter and Saturn, which should produce tidal-rising
cycles with periods of about 10, 12 and 60 years. Jose (1965) noted
that the barycentric motion of the Sun presents a 178.7 year periodicity
and that the rate of change of the Sun's orbital angular momentum
can be correlated to both the 11-year sunspot cycle and
the 22-year magnetic dipole inversion and sunspot polarity Hale cycle.
Bigg (1967) found that the daily sunspot number for the years 1850-1960
presents a consistent periodicity at the sidereal period of Mercury,
which is partially modulated by the position of Venus, Earth and Jupiter.
In a set of studies it has been noted a link between aurora occurrences
and planetary cycles \citep{Charvatova88,scafett2011} and a coherence
between solar motion and solar variability at multiple frequencies
ranging from a 60-year periodicity, which is related to the great
conjunction period of Jupiter and Saturn, to 2400 year \citep{Charvatova90,Charvatova20,Scafetta}.
Landscheidt (1988, 1999) noted that solar rotation and extremes in
sunspot cycle are correlated to solar motion and that this correlation
had very small probabilities, $\sim$ 0.001, of being accidental.
Further evidence of a link involving spin-orbit coupling between the
Sun and the jovian planets has been recently found \citep{Juckett,Wilson}.
Ogurtsov et al. (2002) and Komitov (2009) showed that millennial sunspot
and cosmogenic isotope records such as $^{14}$C and $^{10}$Be are
characterized by major cycles such as at about 45, 60, 85, 128 and 205
years. These cycles can be easily associated to some combination of
planetary cycles. For example: $\sim$45-year is the synodic period
of Jupiter and Uranus; $\sim$60 year is the great conjunction cycle
of Jupiter and Saturn (which is made of three J/S conjunction periods);
$\sim$85-year is the 1/7 resonance of Jupiter and Uranus; and $\sim$205
year is the beat resonance between the 60-year and the 85-year cycles.

\begin{figure}[t]
\begin{centering}
\includegraphics[angle=-90,width=1\textwidth]{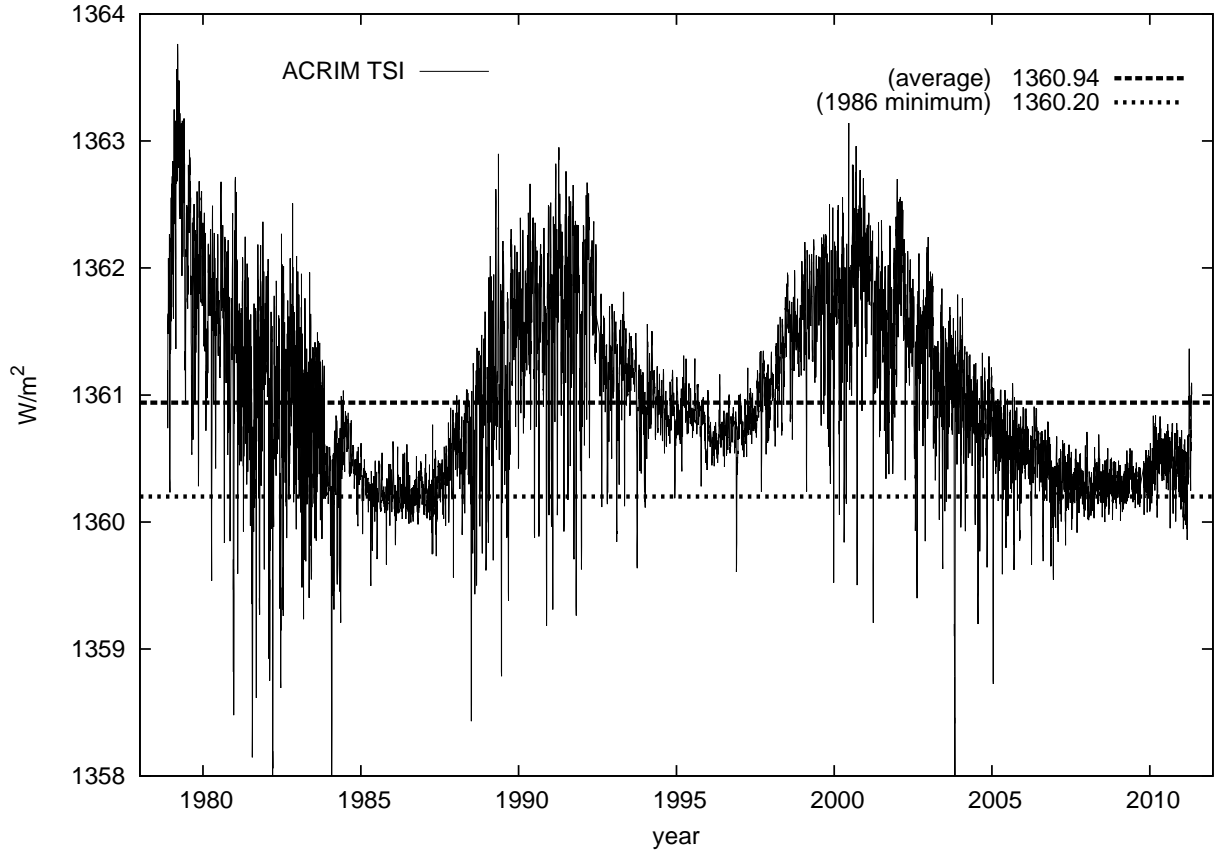}
\par\end{centering}

\caption{ACRIM total solar irradiance (TSI) satellite composite at 1 AU. The
average value is 1360.94 $W/m^{2}$, which implies an average solar
luminosity of $L_{S}=3.827\cdot10^{26}$ W. (Data from http://acrim.com)}
\end{figure}

Twenty-five of the 38 largest known solar flares were observed
to start when one or more tide-producing planets (Mercury, Venus,
Earth, and Jupiter) were either nearly above the event positions ($<10^{o}$
longitude) or at the opposing side of the Sun \citep{Hung}. Active
solar regions are almost symmetrical with respect to the solar equator
and are limited to the medium and low latitudes probably because of
the influence of the planetary tidal force that are stronger at the
medium and low latitudes \citep{Takahashi2}. Actual TSI satellite
observations \citep{Willson}, in particular during the last solar
maximum 1999-2004, present very large quasi annual and sub-annual
cycles in phase with inner planet orbits: for example, Figure 1 shows
the ACRIM TSI composite where these large annual cycles from 1999
to 2004 are macroscopic. These annual cycles may emerge in solar records
only during specific periods as the result of some resonance effect.

A detailed study of the cycles of the Earth's climate and of the cycles
of the speed of the barycentric motion of the Sun has found that the
two systems are strongly coherent at multiple frequencies ranging
from 5 to 100 years \citep{Scafetta,scafett2011}. In particular,
cycles at about 10.4, 20, 30 and 60 years, which are mostly related
to Jupiter and Saturn orbits, are clearly seen in both climate and
solar barycentric motion. This finding would suggest that the planets
are driving solar variability that then drives the Earth's climate.
Note that the 10.4 year cycle seen in the Earth's temperature record
is related to a multiple-planetary alignment, as also found by researchers
that have critiqued the possibility of a planetary-tidal theory \citep{Okal}.
This cycle is likely dominated by the $\sim$10-year spring tide
period of Jupiter and Saturn, which is half of their $\sim$20-year
synodic period.

In the following subsections, we show that the three sub-systems Jupiter-Saturn,
Venus-Earth-Jupiter and Mercury-Venus produce major resonances that
are centered on the 11-year Schwabe solar cycle. These results suggest
that it is unlikely that the Sun is oscillating around the Schwabe's
frequency band (9-13 year periods) just by coincidence. Empirical evidences clearly suggest that
solar dynamics is approximately synchronized to planetary motion.

\subsection{The 10-12 year Jupiter and Saturn cycles in the sunspot number record.}

\begin{figure}[t]
\begin{centering}
\includegraphics[angle=-90,width=1\textwidth]{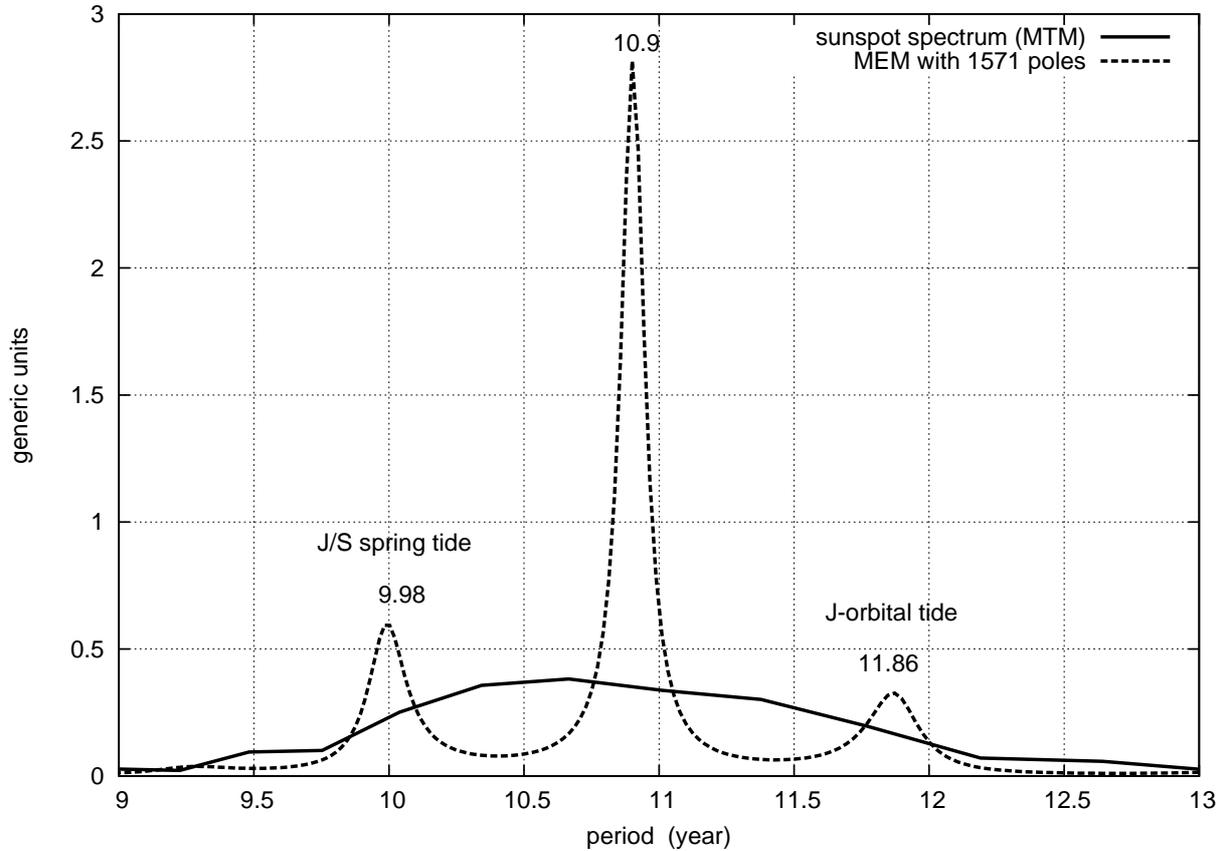}
\par\end{centering}

\caption{Power spectrum analysis of the 11-year Schwabe's solar cycle as deduced
from the sunspot number record monthly sampled from 01/1749 to 12/2010
(3144 data) (http://sidc.oma.be/sunspot-data/). Multi taper method
(MTM) and maximum entropy method (MEM) with 1572 pole order are used
(Ghil et al., 2002). In the latter case a reasonable error of about
$\pm0.1$ year related to the monthly sampling can be assumed. The
three peaks have a 99\% confidence level against red noise background.
See \citet{Scafetta200} for details. }
\end{figure}

\citet{Scafetta200} has recently shown that the sunspot number record
presents a wide 9-13 year Schwabe's frequency band made of three frequencies.
These are compatible with Jupiter/Saturn spring tidal period of 9.93
years (note that ephemeris calculations of the orbits of Jupiter and
Saturn since 1749 give a J/S spring tidal cycle of $9.93\pm0.5$ year)
with a central 10.87$\pm0.1$ year period, which is likely related
to a central 11-year solar dynamo cycle, and with the period of Jupiter
at 11.86 years. These patterns suggest that the 11-year solar cycle
may be partially induced by some solar resonance driving the solar
dynamo that amplifies the beat average cycle (10.81-10.90 years) induced
by the two major tides produced by the combined system of Jupiter-Saturn
($\sim$9.93-year cycle) and by Jupiter's orbit ($\sim$11.86-year
cycle), plus a possible small dynamical correction induced by other
tides as explained below. Figure 2 shows the three frequency peaks
of the sunspot number record. According to the power spectrum evaluation,
it may be possible to associate at least 0.1-0.5 $W/m^{2}$ variability
to the planetary tides.

The existence of three  cycles imply additional beat cycles.
For example, there exists a beat at

\begin{equation}
T=\left(\frac{1}{T_{JS}}-\frac{1}{T_{J}}\right)^{-1}=61\pm2~year\label{beatcc}
\end{equation}
This quasi-sexagesimal cycle is found in millennial cosmogenic isotope
records such as $^{14}$C and $^{10}$Be and in the aurora records
\citep{Charvatova88,Ogurtsov,Komitov}, and in numerous terrestrial
climate records \citep{Scafetta,Scafetta200}. Other beats occur at
about 115, 130 and 983 years. \citet{Scafetta200} extensively studied the above findings and showed that these frequencies can be
used to explain all known major solar patterns, which include a sufficiently
correct timing of the 11-year Schwabe solar cycle and of the known
grand solar minima during the last millennium known as the Oort, Wolf,
Spörer, Maunder and Dalton minima, plus the emergence and the timing
of observed quasi-millennial solar and climate cycles.

\subsection{The Venus-Earth-Jupiter 11.07-year alignment cycle.}

Venus, Earth and Jupiter are the three major tidal planets (see Section
3 and Table 2 for the estimated values of the tidal elongation induced
on the Sun by each planet). It has been found that Venus, Earth and
Jupiter tend to be mostly aligned every 11 years \citep{Bendandi,Takahashi, Wood, Hung}.
Indeed, when the combined alignment of Venus, Earth and Jupiter is
taken into account, the configuration E-V-Sun-J or Sun-V-E-J repeats
with a period of approximately 22 years, and the correspondent tidal
period is half of it. This period can be calculated  using the following
resonance formula:

\begin{equation}
P_{VEJ}=\frac{1}{2}\left(\frac{3}{P_{V}}-\frac{5}{P_{E}}+\frac{2}{P_{J}}\right)^{-1}=4043~day=11.07~year,\label{VEJ1}
\end{equation}
where $P_{V}=224.701$ days, $P_{E}=365.256$ days and $P_{J}=4332.589$
days are the sidereal periods of Venus, Earth and Jupiter, respectively.

It is possible to define a very simple multi-planetary alignment index
for Venus, Earth and Jupiter \citep{Hung}. Let us assume that the
Sun is in the center of a system and that three planets are orbiting
the Sun. An alignment index $I_{ij}$ between planet $i$ and planet
$j$ can be defined as:

\begin{figure}[t]
\begin{centering}
\includegraphics[angle=-90,width=1\textwidth]{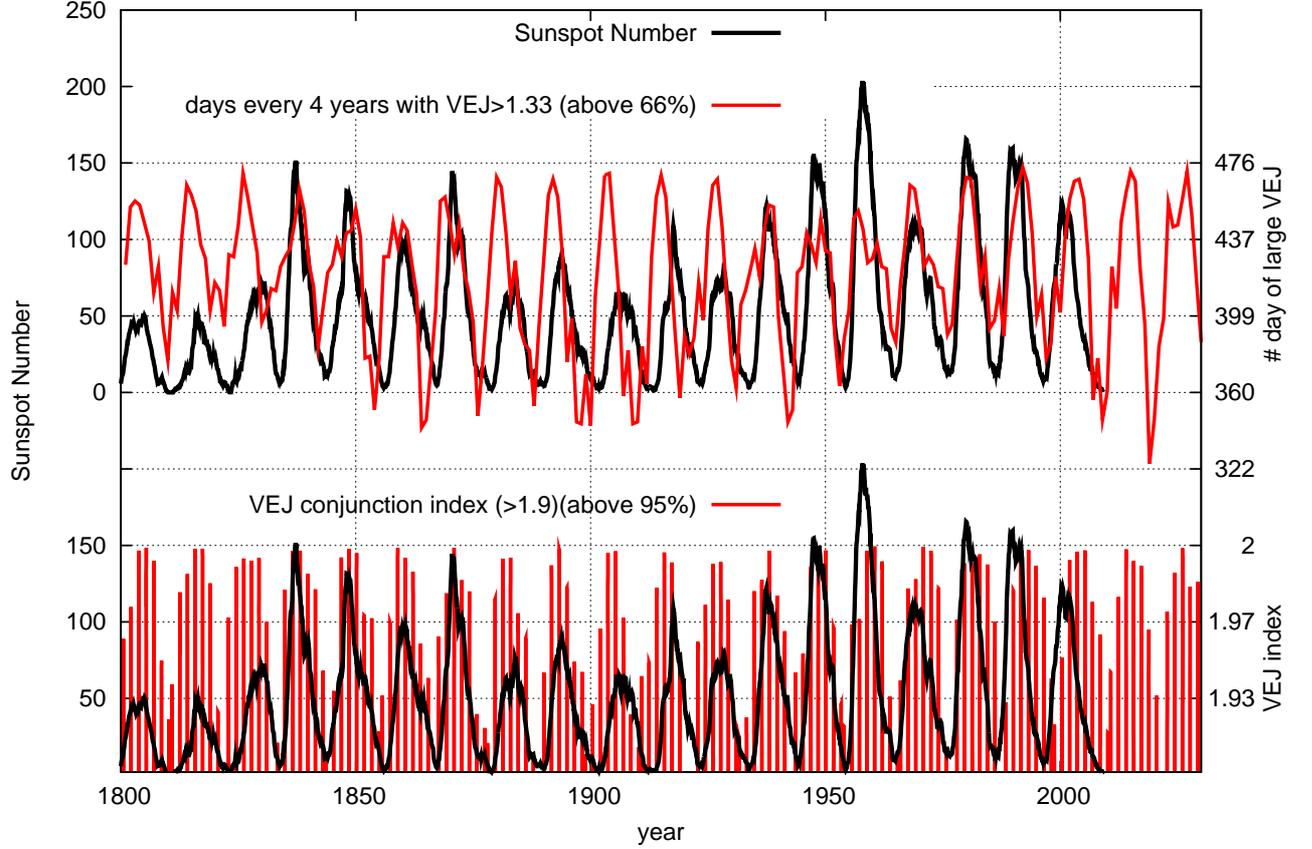}
\par\end{centering}

\caption{(Top) the sunspot number record (black) is compared against the number
of days (every four years) (red) when the alignment index $I_{VEJ}>66\%$.
(Bottom) the sunspot number record (black) is compared against the
most aligned days with $I_{VEJ}>95\%$ (red).}
\end{figure}

\begin{equation}
I_{ij}=|\cos(\Theta_{ij})|,\label{eq1}
\end{equation}
where $\Theta_{ij}$ is the angle between the positions of the two
planets relative to the solar center. Eq. \ref{eq1} indicates that
when the two planets are aligned ($\Theta_{ij}=0$ or $\Theta_{ij}=\pi$)
the alignment index gives the largest value because these two positions
imply the largest combined tidal elongation on the Sun, which is known
as the \emph{spring tide}. This is the synodic/opposition tidal period
(which is half of the synodic period of the two planets because tidal
forces produce two opposite tides). When $\Theta_{ij}=\pi/2$ the
index gives the lowest value because at right angles the tides of
the two planets tend to cancel each other, which is known as the \emph{neap
tide}. In the case of a system made of Venus, Earth and Jupiter, there
are three correspondent alignment indexes:

\begin{eqnarray}
I_{V} & = & |\cos(\Theta_{VE})|+|\cos(\Theta_{VJ})|\\
I_{E} & = & |\cos(\Theta_{EV})|+|\cos(\Theta_{EJ})|\\
I_{J} & = & |\cos(\Theta_{JV})|+|\cos(\Theta_{JE})|.
\end{eqnarray}
Finally, we can define a combined alignment index $I_{VEJ}$ for the
three planets as

\begin{equation}
I_{VEJ}=smallest~among~(I_{V},I_{E},I_{J}).\label{eq2}
\end{equation}
The above index $I_{VEJ}$ has a value ranging from 0 to 2. It can
be carefully calculated by using the daily coordinates of Venus, Earth
and Jupiter relative to the Sun, which can be obtained by using the
high-precision JPL HORIZONS ephemerides system ({http://ssd.jpl.nasa.gov/}).

Figure 3 shows two alternative comparisons between the annual average
of sunspot number since 1800 and two alternative indexes obtained
with the alignment index $I_{VEJ}$. In the top of the figure the
sunspot number record is compared against the number of days (every
four years) that present an alignment index $I_{VEJ}>66\%$; in the
bottom of the figure the same sunspot number record is compared against
the most aligned days with $I_{VEJ}>95\%$ (or $I_{VEJ}>1.9$).

The figure shows that there exists a very good coherence between the
sunspot number record and the two indexes obtained by using the alignment
index $I_{VEJ}$. In particular, note the relatively good phase synchronization.
The two records present 19 full cycles for the covered period 1800-2010
with an average period of about 11.05 year per cycle, which is compatible
with Eq. \ref{VEJ1}. Note that the average of the observed 23 sunspot
cycle lengths since 1750 is about 11.06 years.

\begin{figure}[t]
\begin{centering}
\includegraphics[angle=-90,width=1\textwidth]{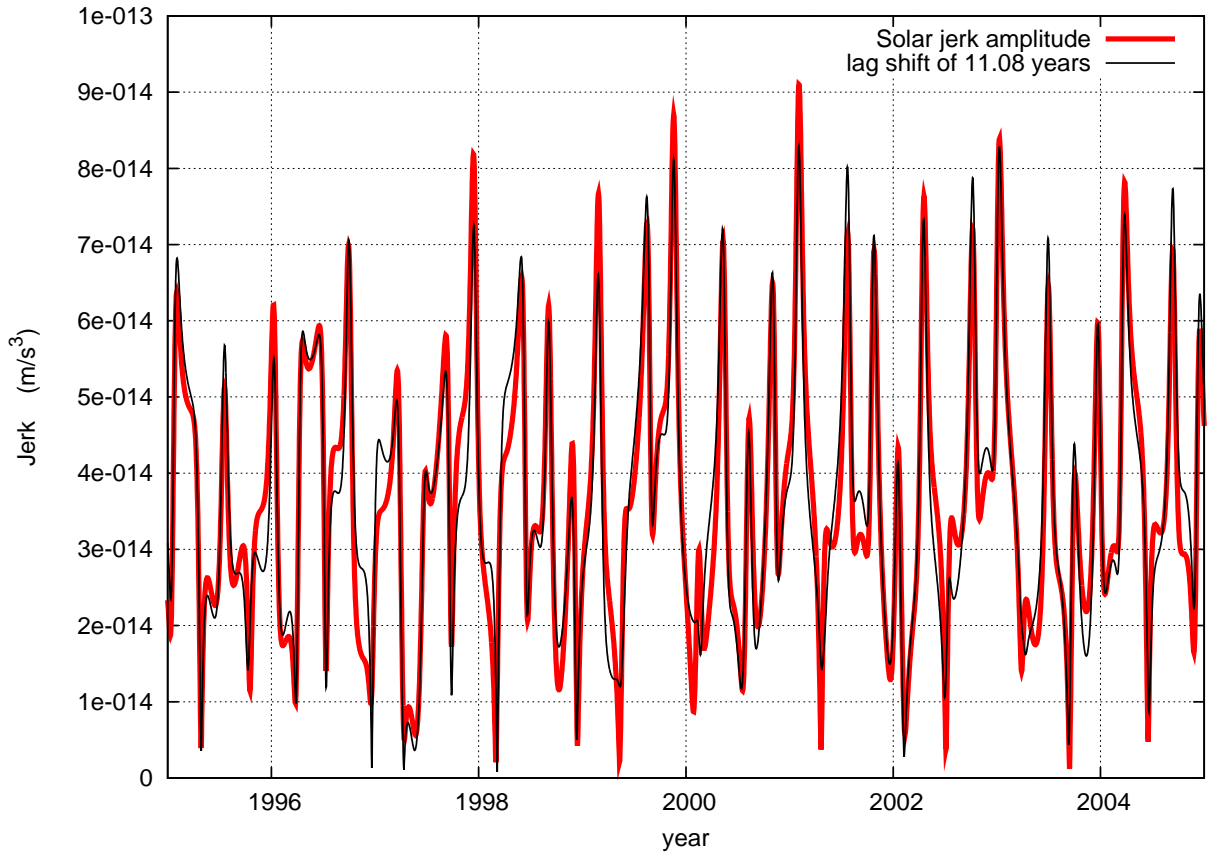}
\par\end{centering}

\caption{Solar jerk amplitude potted against itself with a time-lag of 11.08
years. Note the extremely good correlation that demonstrates the existence
of a 11.08 year recurrence in the solar jerk function. The plotted
periods are 1995-2005 (gray) and 2006.08-2016.08 (black).}
\end{figure}

\subsection{The 11.08-year solar jerk-shock vector cycle.}

The jerk vector is the time derivative of the acceleration vector
\citep{Schot} and it is commonly used in engineering as a measure
of the shocks and stresses felt by an object subject to a varying
acceleration. Jerk functions are associated with dissipative chaotic
flow mechanisms \citep{Sprott}, and changes in the acceleration of
vortices are associated with the generation of flow noises \citep{Schot}.
Jerk stresses may contribute to solar variations \citep{Wood65}.
The Sun's acceleration vector is given by

\begin{equation}
\vec{a}_{S}(t)=\sum_{P=1}^{8}\frac{G~m_{P}~\vec{R}_{SP}(t)}{R_{SP}^{3}(t)},\label{201}
\end{equation}
where $\vec{R}_{SP}(t)$ is the vector position of a planet from the
Sun. The sum is extended to the eight planets of the solar system
(Mercury, Venus, Earth, Mars, Jupiter, Saturn, Uranus and Neptune).
The Sun's jerk vector is given by

\begin{equation}
\vec{J}_{S}(t)=\dot{\vec{a}}_{S}(t)=\sum_{P=1}^{8}\frac{G~m_{p}~V_{SP}(t)}{R_{SP}^{3}(t)}\left(\frac{\vec{V}_{SP}(t)}{V_{SP}(t)}-3~\frac{\vec{R}_{SP}(t)}{R_{SP}(t)}\right),\label{202}
\end{equation}
where $\vec{V}_{SP}(t)$ is the velocity of the planet P relative
to the Sun.

\begin{figure}
\begin{centering}
\includegraphics[width=0.9\textwidth]{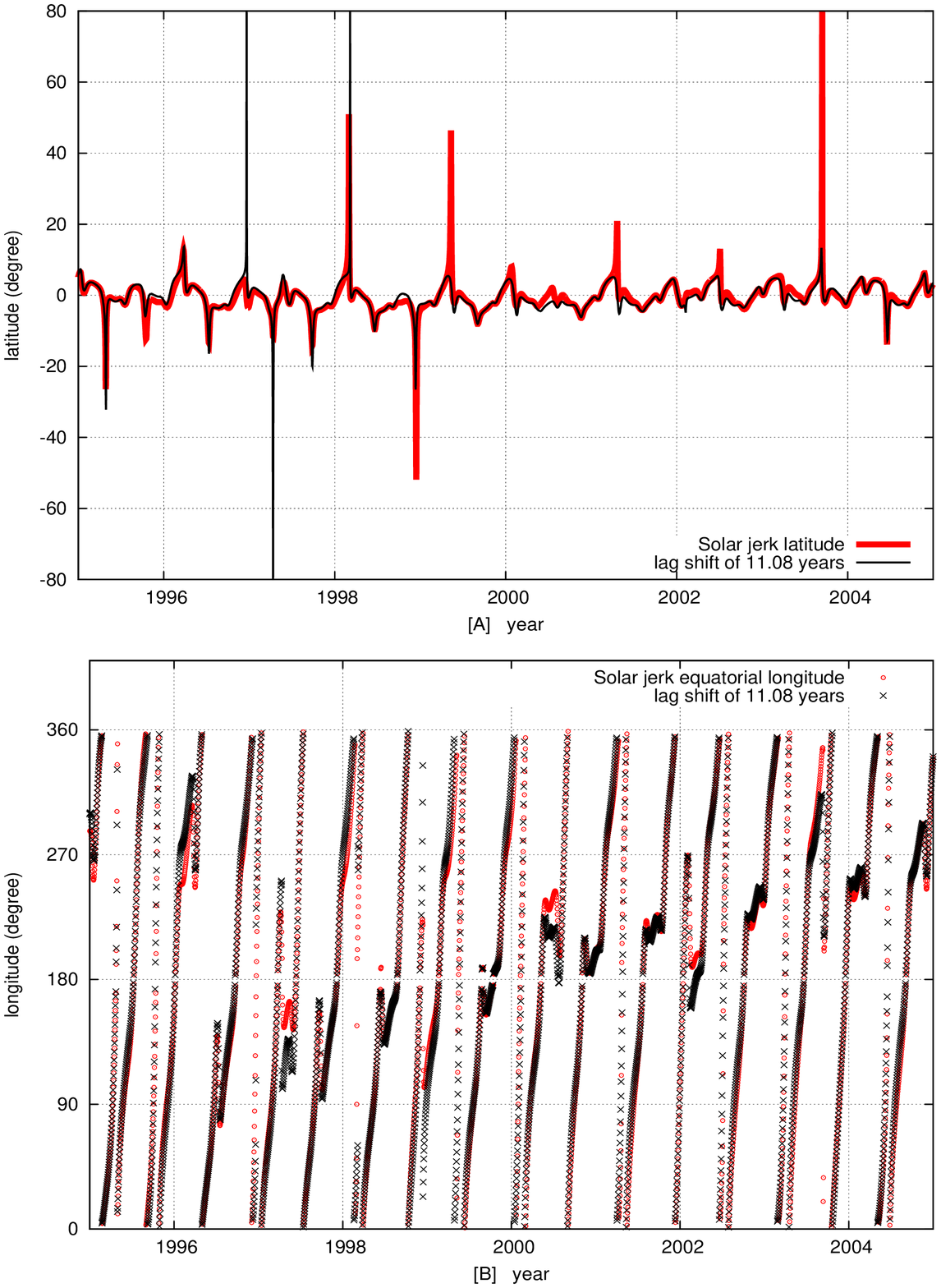}
\par\end{centering}

\caption{{[}A{]} Solar jerk latitude and {[}B{]} longitude potted against themselves
with a time-lag of 11.08 years. Note the extremely good correlation
that demonstrates the existence of a 11.08 year recurrence in the
solar jerk vector rotation. The plotted periods are 1995-2005 and
2006.08-2016.08 as in Figure 4.}
\end{figure}

Eq. \ref{202} is also closely related to the tidal acceleration induced
by the planets, which is proportional to the factor $G~m_{p}/R_{SP}^{3}(t)$.
The jerk vector weights the tidal acceleration with the velocity of
each planet. If two planets induce the same tidal acceleration, the
one that moves faster produces the greatest jerk vector on the Sun,
and, thus, it induces the greatest jerk-stress and torsion on the
solar plasma.

Figure 4 shows the magnitude of the solar jerk vector given by Eq.
\ref{202}. Fast fluctuations due to the movement of the inner planets
are seen. The autocorrelation is maximum at 11.08-year lag. For example,
the figure shows an extremely good correlation between the ten year
period 1995-2005 and the ten year period 2006.08-2016.08.

Figure 5 shows the angular evolution of the solar jerk latitude and
longitude. The jerk vector rotates irregularly around the Sun following
the planetary orbits. Latitude and longitude are plotted against themselves
with a time-lag of 11.08-year, which again reveals a 11.08-year periodic
cycle in the solar jerk vector.

By assuming circular orbits and using the third Kepler's law, the
speed of each planet, as seen from the Sun, is given by

\begin{equation}
V_{SP}=\sqrt{\frac{G~M_{S}}{R_{SP}}}~.\label{203}
\end{equation}
Thus, the amplitude of the solar jerk vectors corresponding to each
planet is

\begin{equation}
J_{P}=\frac{G^{1.5}~m_{p}~\sqrt{10M_{S}}}{R_{SP}^{3.5}}~.\label{204}
\end{equation}
The vector is oriented on the planetary orbital plane at an angle

\begin{equation}
\beta=atan\left(-\frac{1}{3}\right)=+161.5651^{o}~.\label{205}
\end{equation}
from the sun-planet distance vector. Table 1 shows the values of $J_{P}$
for each planet. Venus and Mercury dominate this variable. Earth and
Jupiter produce weaker and almost compatible jerk vectors.

\begin{table}
\begin{centering}
\begin{tabular}{cccc}
\hline
Planet & mass  & semi-major  & $J_{P}$ \tabularnewline
 & (kg)  & axis (m)  & (m/s$^{3}$) \tabularnewline
\hline
Me  & 3.30E+023  & 5.79E+010  & 1.72E-14 \tabularnewline
\hline
Ve  & 4.87E+024  & 1.08E+011  & 2.84E-14 \tabularnewline
\hline
Ea  & 5.97E+024  & 1.50E+011  & 1.12E-14 \tabularnewline
\hline
Ma  & 6.42E+023  & 2.28E+011  & 2.76E-16 \tabularnewline
\hline
Ju  & 1.90E+027  & 7.79E+011  & 1.11E-14 \tabularnewline
\hline
Sa  & 5.69E+026  & 1.43E+012  & 3.92E-16 \tabularnewline
\hline
Ur  & 8.68E+025  & 2.88E+012  & 5.22E-18 \tabularnewline
\hline
Ne  & 1.02E+026  & 4.50E+012  & 1.28E-18 \tabularnewline
\hline
\end{tabular}
\par\end{centering}

\caption{Amplitude of the jerk vector on the Sun for each planet: see Eq. \ref{204}. }

\label{tb4}
\end{table}

Note that Mercury and Venus orbital combination repeats almost every
11.08 years. In fact, $46P_{M}=11.086$ year and $18P_{V}=11.07$
year, where $P_{M}=0.241$ year and $P_{V}=0.615$ year are the sidereal
periods of Mercury and Venus, respectively. Indeed, Mercury and Venus's
orbits repeat every 5.54 years, but the 11.08 year periodicity would
better synchronize with the orbits of the Earth and with the tidal
cycles of the sub-system Jupiter-Saturn. This resonance also implies
that the tidal cycle produced by Mercury and Venus repeats every 5.54
and 11.08 years.

The 11.08 year recurrent pattern in the orbit of Mercury
and Venus as well as the 11.07 year cycle in the alignment of Venus,
Earth and Jupiter, which was discussed above (Eq. \ref{VEJ1}), are
particularly interesting because the mean solar cycle length since
1750 is estimated to be 11.06 years (in 1900 Brown found 11.1-years),
which is extremely close to such planetary resonance periods.
Perhaps these frequencies slightly shift the central Schawabe
frequency from 10.81-year to 10.87-year \citep{Scafetta200}.

\section{Estimation of the planetary tidal heating and its nuclear fusion amplification
factor in the solar core}

 In this section we attempt to roughly estimate the luminosity
variation that planetary tides can hypothetically induce, and we compare
our estimate against the observed roughly 1 $W/m^{2}$ 11-year TSI
cycle variation at 1 AU, as seen in Figure 1.

As explained in the Introduction, the idea that we propose is that
nuclear fusion mechanisms in the solar core greatly amplify the gravitational
energy released by tidal forcings in the core. Only by means of a
huge nuclear fusion amplification mechanism may planetary tides produce irradiance
output oscillations with a sufficient magnitude relative to a background
noise to influence solar dynamo processes. In fact, it is very unlikely that dynamical mechanisms
such as synchronization and resonance effects can  be
activated without an initial strong nuclear fusion amplification of the original
gravitational tidal energetic signal, which by alone would be far
too weak compared with competing energetic random flows present in
the solar convective zone, as also indirectly argued in \citet{Jager}.
Here we explain how to calculate at first approximation a nuclear
magnification factor converting released gravitational potential energy
into solar luminosity by postulating the existence of a physical link
between the solar luminosity rate and the potential energy associated
to the mass lost by nuclear fusion in an unit of time. The needed equation
can be deduced from  the well-known \emph{mass-luminosity relation} \citep{Duric}.

\subsection{Basic theory of tides.}

The near-surface rotation rate of the Sun is observed to be larger
at the equator (about 25 days) and to decrease as latitude increases
(up to about 36 days at the poles) \citep{Beck}: current measurements
are commonly expressed by a function of the form of $A+B\sin^{2}(\theta)+C\sin^{4}(\theta)$,
where $\theta$ is the latitude of the Sun. The radiative zone of
the Sun is found to rotate roughly uniformly with a period of about
26-27 days, while the core may rotate with a compatible rate \citep{Thompson}.
Because herein we are only interested in finding a rough estimate
of the effects of the tides on the Sun, for simplicity, we assume
that the solar interior is rotating uniformly with a period of 26.5
days.

The basic tidal potential height equation is

\begin{equation}
H_{t}(\alpha,r)=\frac{3}{2}~H_{tm}(r)\left[\cos^{2}(\alpha)-\frac{1}{3}\right],\label{eq31}
\end{equation}
where $\alpha$ is the angle, relative to the solar center, between
the position of a planet and the position of a location $A$ on the
Sun, and $H_{tm}(r)$ is the high of the tide at $\alpha=0$ and distance
$r$ from the center of the Sun \citep{Lamb}. The value of $H_{tm}(r)$
depends on several variables such as the distance between the Sun
and the planet, the distance $r$ between the center of the Sun and
the location $A$, and the mass of the Sun within the radius $r<R_{S}$.

Because the planetary tides are very small we assume the Sun to be
perfectly spherical and use the average solar radius, $R_{S}=6.956\cdot10^{8}$
m. According to such approximation, the difference between the highest
(at $\alpha=0$) and the lowest (at $\alpha=\pi/2$) tidal points,
which is called \emph{tidal elongation}, is given by \citep{Taylor}

\begin{equation}
T_{E}(r)=\frac{3}{2}~H_{tm}(r)=\frac{3}{2}\frac{m_{P}}{m_{S}(r)}\frac{r^{4}}{R_{SP}^{3}},\label{eq32}
\end{equation}
where $m_{P}$ is the mass of a planet $P$, $m_{S}(r)$ is the mass
of the Sun included within the radius $r\leq R_{S}$ and $R_{SP}$
is the distance between the Sun and the planet $P$.

The initial factor 3/2 in Eq. \ref{eq32} could be substituted with
other \emph{Love} numbers, which depend on the physical properties
of the body deformed by the tides. The adoption of the Love factor
3/2, which refers to the tidal deformation of a small fluid shell
above a rigid sphere, approximately like the case of the ocean tides
above the Earth's more rigid crust, gives a lower limit. However,
a Love factor 15/4 should be used if the perturbed Sun assumes the
form of an uniform spheroid, which refers to an uniform fluid ball
\citep{Fitzpatrick}. The Love number for the Sun may be between the
two situations. In the following we use the factor 3/2 in the equations,
but in the figures and in the tables we will use a double index referring
to both Love numbers, 3/2 and 15/4. Note that the ratio between the
two Love numbers is 2.5, so it is easy to convert one result into
the other.

\begin{table}
\begin{centering}
\begin{tabular}{cccccccc}
\hline
Planet  & mass  & semi-major  & perihelion  & aphelion  & tidal  & diff. tidal  & S. rot. \tabularnewline
 & (kg)  & axis (m)  & (m)  & (m)  & elong. (m)  & elong. (m)  & (days) \tabularnewline
\hline
Me  & 3.30E23  & 5.79E10  & 4.60E10  & 6.98E10  & 3.0E-4 (7.5E-4)  & 4.3E-4 (1.1E-3)  & 37.92 \tabularnewline
\hline
Ve  & 4.87E24  & 1.08E11  & 1.08E11  & 1.09E11  & 6.8E-4 (1.7E-3)  & 2.6E-5 (6.6E-5)  & 30.04 \tabularnewline
\hline
Ea  & 5.97E24  & 1.50E11  & 1.47E11  & 1.52E11  & 3.2E-4 (7.9E-4)  & 3.2E-5 (7.9E-5)  & 28.57 \tabularnewline
\hline
Ma  & 6.42E23  & 2.28E11  & 2.07E11  & 2.49E11  & 9.6E-6 (2.4E-5)  & 5.5E-6 (1.4E-5)  & 27.56 \tabularnewline
\hline
Ju  & 1.90E27  & 7.79E11  & 7.41E11  & 8.17E11  & 7.1E-4 (1.8E-3)  & 2.1E-4 (5.2E-4)  & 26.66 \tabularnewline
\hline
Sa  & 5.69E26  & 1.43E12  & 1.35E12  & 1.51E12  & 3.4E-5 (8.5E-5)  & 1.2E-5 (2.9E-5)  & 26.57 \tabularnewline
\hline
Ur  & 8.68E25  & 2.88E12  & 2.75E12  & 3.00E12  & 6.4E-7 (1.6E-6)  & 1.7E-7 (4.3E-7)  & 26.52 \tabularnewline
\hline
Ne  & 1.02E26  & 4.50E12  & 4.45E12  & 4.55E12  & 2.0E-7 (5.0E-7)  & 1.3E-8 (3.3E-8)  & 26.51 \tabularnewline
\hline
\end{tabular}
\par\end{centering}

\caption{Average theoretical tidal elongation at the solar surface, due to
the eight planets of the solar system (Eq. \ref{eq32}). For the Sun
we use $M_{S}=1.99E30$ kg and $R_{S}=6.96E8$ m. The `diff tidal
elongation' refers to the difference of the tidal elongations calculated
at the perihelion and the aphelion, respectively. The last column
reports the solar rotation as seen from each planet using a sidereal
solar rotation of 26.5 days: longer values imply slower vertical tidal
movement and less tidal work in unit of time. The tidal elongations
are calculated with the Love numbers 3/2 and 15/4, the latter in parentheses. }

\label{tb1}
\end{table}

For example, if we use the Love number 3/2, given the solar total
mass ($M_{S}=1.989\cdot10^{30}$ kg), on the Sun's surface Jupiter
and Venus induce tidal elongations of about 0.68-0.71 mm, and Earth
and Mercury induce tidal elongations of about 0.30-0.31 mm: see Table
2. The other planets induce even smaller tidal elongations. The highest
combined tides are produced during the synodic/opposition periods
among the planets: the \emph{spring tides}. For example, a Jupiter-Venus
spring tidal elongation is 1.4 mm on average. The minimum combined
tides occur when right angles among the planets are formed: the \emph{neap
tides}. Table 2 also reports the difference of the tidal elongations
calculated at the perihelion and at the aphelion of each planet. In
this case Mercury produces the largest maximum orbital tidal variation
(about 0.43 mm), Jupiter follows (about 0.21 mm), and Earth, Venus,
Saturn and Mars have orbital tidal variation from 0.03 mm to 0.005
mm. These numbers are extremely small and would still be tiny if we
adopt the Love number 15/4 and multiply them by 2.5: see Table 2.
No planetary physical effect on the Sun would be measurable unless
some huge internal amplification energetic mechanism exists. In fact,
it is highly unlikely that internal dynamical synchronization and
resonance mechanisms alone could be activated by such tiny tides because
the gravitational tidal effect alone would be completely covered by
dynamical noise in the convective zone, and the noise would efficiently
disrupt the signal before it may become measurable. In fact, solar
dynamics needs first to synchronize to the planetary frequencies,
but if the thermal and dynamical noise is far too strong relative
to the original tidal signal, numerous phase slips would occur, no
synchronization would take place \citep{Pikovsky} and resonance mechanisms
would not be activated.

By comparison, the tidal elongations induced by the Moon ($R_{EM}=3.844\cdot10^{8}$
m and $M_{M}=7.348\cdot10^{22}$ kg) and by the Sun on the Earth ($R_{E}=6.37\cdot10^{6}$
m) are much larger: 0.54 m and 0.25 m, respectively, which are measurable
quantities, and dynamical resonances may increase the signal up to
a factor 10 in certain locations. Thus, we may expect that synchronization
and dynamical resonance mechanisms alone could eventually increase
the planetary tidal signal on the Sun by one or two orders of magnitude
at most, but nothing would be observed if the energetic gap would
be of six or more order of magnitudes!

Indeed, using  Newtonian classical physics alone, the tidal elongations induced by the planets on the Sun appear
to be extremely small and have discouraged researchers from believing
that planetary gravity could influence solar activity. However, we
believe that the physical quantity of interest is not the tidal work
released to the Sun alone, but the solar nuclear fusion feedback response
to it that has the potentiality of greatly amplifying the gravitational
energetic tidal signal. In the following subsections we  roughly estimate
the tidal power dissipated inside the Sun and the solar nuclear positive feedback response to it.

\subsection{Estimation of the planetary tidal gravitational power dissipated
in the Sun.}

In the following we estimate the total maximum power that the planetary
tides may hypothetically release to the Sun. At each location of the
Sun with spherical coordinates ($r,\theta,\phi$), in a given time
interval $\Delta t$, a given mass

\begin{equation}
\texttt{d}m(r,\theta,\phi)=\rho(r)r^{2}\sin(\theta)\texttt{d}r\texttt{d}\theta\texttt{d}\phi\label{}
\end{equation}
is potentially vertically moved (up or down) by $|\Delta h(r,\theta,\phi)|$
by the tidal forces. The function $\rho(r)$ is the density of the
Sun at the distance $r$ (we assume a perfect spherical symmetry).

Herein we do not take into account the effects of the horizontal tides
because the maximum energy involved in the process is determined by
the vertical oscillation. However, the movement of the horizontal
tides is also important because, together with the vertical tides,
it can induce a mixing of solar masses that can carry fresh fuel to
deeper levels \citep{Wolff} and increase nuclear fusion rate. The
absolute value in the above equation is used because a tide, by working
against the solar inertia to gravitational deformation, would heat
the solar masses both when it rises and when it falls.

The total power related to the work associated to the vertical movement
of the tide in a given location is given by

\begin{eqnarray}
\Delta P(r,\theta,\phi) & = & \frac{\Delta W(r,\theta,\phi)}{\Delta t}\nonumber \\
 & = & \frac{|\Delta h(r,\theta,\phi)|~g(r)~\rho(r)~r^{2}~\sin(\theta)~\texttt{d}r\texttt{d}\theta\texttt{d}\phi}{\Delta t}
\end{eqnarray}
where $g(r)$ is the acceleration of gravity of the Sun at the distance
$r$. The total power is given by a spherical integration on the entire
Sun because tides affect also the inner part of the Sun, not just
the solar surface or  the tachocline. However, only a very small fraction of this power
is released to the star. Thus, we write

\begin{equation}
P=\frac{1}{Q}\int_{r=0}^{R_{S}}\int_{\theta=0}^{\pi}\int_{\phi=0}^{2\pi}\Delta P(r,\theta,\phi).\label{eq40}
\end{equation}
The parameter $Q^{-1}$ is called \emph{effective tidal dissipation
function}, and it is defined as the energy lost during one complete
cycle over the maximum energy stored in the tidal distortion: $Q^{-1}=\frac{1}{2\pi E_{o}}\oint(-\frac{dE}{dt})dt$
\citep{Goldreich,Jackson}.

The exact determination of the solar $Q$ factor goes beyond the purpose
of this work, and for the Sun it is unknown. This factor is usually
estimated indirectly through an analysis of the temporal evolution
of orbital parameters in binary solar systems. In fact, it is known
that the tidal torques transfer angular momentum and energy from the
planet's rotation into the satellite's orbital revolution. In the
case of giant planets such as Jupiter, it is possible to study the
orbits of their moons, and it is found a value of $Q\approx10^{5}$
\citep{Goldreich}. In the case of stars, the same methodology can
be applied to the orbital evolution of binary systems \citep{Tassoul},
and for solar-like stars a reasonable range is $5\cdot10^{5}\lessapprox Q\lessapprox2\cdot10^{6}$,
according to the empirical results depicted in figure 1 in \citet{Ogilvie}:
see also \citet{Meibom}. However, a very large uncertainty exists
about the parameter $Q$ also because it is a function that depends
on several variables such as the frequency and strength of the tides, and largely changes from
star to star. Herein we adopt a constant value for all planetary tides:

\begin{equation}
Q=10^{6},\label{Q41}
\end{equation}
observing that there may be an error of one order of magnitude and that Q may be larger (smaller) for tides with larger (smaller) amplitude and frequencies: we will address these important corrections in another work. The
large value of $Q$ indicates that only a tiny fraction of the tidal
associated power can be transferred to the Sun because of the solar interior
rigidity.

The acceleration of gravity at a certain position $r$ is given by

\begin{equation}
g(r)=\frac{G~m_{S}(r)}{r^{2}}.\label{eq41}
\end{equation}
where the solar mass $m_{S}(r)$ within the radius $r\leq R_{S}$
is given by

\begin{equation}
m_{S}(r)=4\pi\int_{0}^{r}\rho(r')~r'^{2}~\texttt{d}r'.\label{eq61}
\end{equation}
Figure 6 shows a typical solar density function $\rho(r)$ obtained
from standard solar models \citep{Bahcall}.

\begin{figure}[t]
\begin{centering}
\includegraphics[angle=-90,width=1\textwidth]{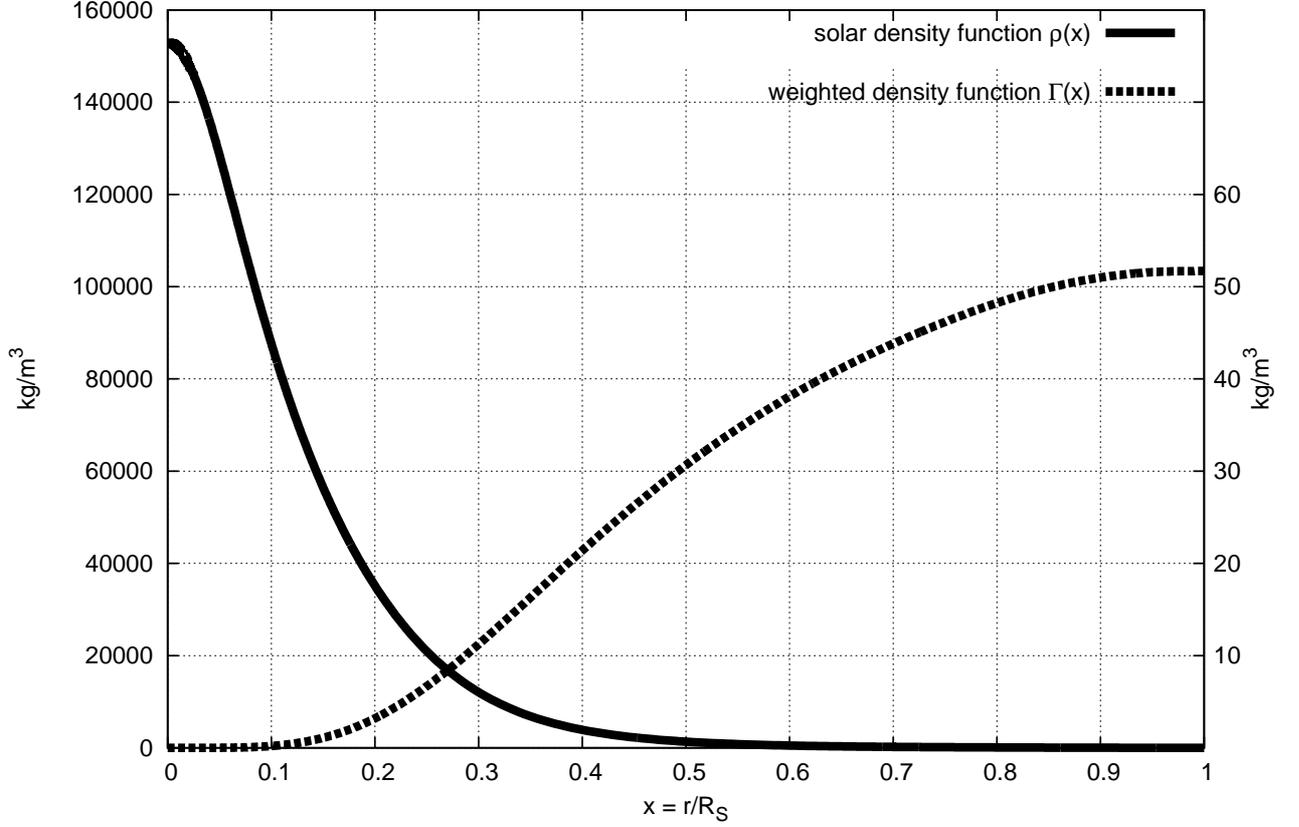}
\par\end{centering}

\caption{A typical standard solar model density function $\rho(r/R_{S})$ (solid
line). The figure depicts the model BS05(OP) (Bahcall et al., 2001,
2005) which is constructed with a traditional understanding of the
current heavy element abundance on the Sun. The weighted solar density
function $\Gamma(r/R_{S})$, Eq. \ref{eq55} (dash line) refers to
the values at the right ordinates. Three values may be of special
interests: at the solar core radius, $\Gamma(0.3)=11.23$ $kg/m^{3}$;
at the solar radiative zone radius, $\Gamma(0.714)=44.5$ $kg/m^{3}$;
and at the solar near-surface radius, $\Gamma(1)=51.7$ $kg/m^{3}$. }
\end{figure}

In the following we consider the tide produced by only one planet
P. The vertical displacement $|\Delta h(r,\theta,\phi)|$ can be calculated
using Eq. \ref{eq31}. If the position of a planet P at a given time
$t$ in Cartesian coordinates is given by ($X_{t},Y_{t},Z_{t}$),
and the position of a point A in the spinning Sun is ($x_{t},y_{t},z_{t}$)
(in solar equatorial coordinate plane), we have

\begin{eqnarray}
\cos(\alpha_{P,t}) & = & \frac{x_{t}X_{t}+y_{t}Y_{t}+z_{t}Z_{t}}{r~R_{SP}(t)}\nonumber \\
 & = & \frac{\sin(\theta)\cos(\phi_{t})X_{t}+\sin(\theta)\sin(\phi_{t})Y_{t}+\cos(\theta)Z_{t}}{R_{SP}(t)},
\end{eqnarray}
where $R_{SP}(t)$ is the distance between the center of the Sun and
the planet at the time $t$, and $\theta$ and $\phi_{t}$ are the
latitude and the longitude of the solar location. The angle $\phi_{t}$
depends on $t$ because of the solar rotation.

Here, for simplicity, we suppose that the Sun is rotating at a constant
angular velocity $\Omega=2\pi/T$, where $T=26.5$ days, as explained
above. The vertical tidal displacement in a time interval $\Delta t$
due to the planet P is calculated using Eqs. \ref{eq31} and \ref{eq32},
and it is given by

\begin{equation}
|\Delta h_{P}(r,\theta,\phi_{t})|=\frac{3~r^{4}~m_{P}}{2~m_{S}(r)}\left|\frac{\cos^{2}(\alpha_{P,t})-\frac{1}{3}}{R_{SP}^{3}(t)}-\frac{\cos^{2}(\alpha_{P,t-\Delta t})-\frac{1}{3}}{R_{SP}^{3}(t-\Delta t)}\right|,\label{eq51}
\end{equation}
where the angles $\alpha_{P,t}$ and $\alpha_{P,t-\Delta t}$ are
calculated using Eq. \ref{eq42} by taking also into account the rotation
of the Sun:

\begin{equation}
\phi_{t-\Delta t}=\phi_{t}-\Omega~\Delta t~.\label{eq52}
\end{equation}
By substituting the above equations in Eq. \ref{eq40}, we find that
the maximum tidal released power within the radius fraction $r/R_{S}$
is given by

\begin{figure}[t]
\begin{centering}
\includegraphics[angle=-90,width=1\textwidth]{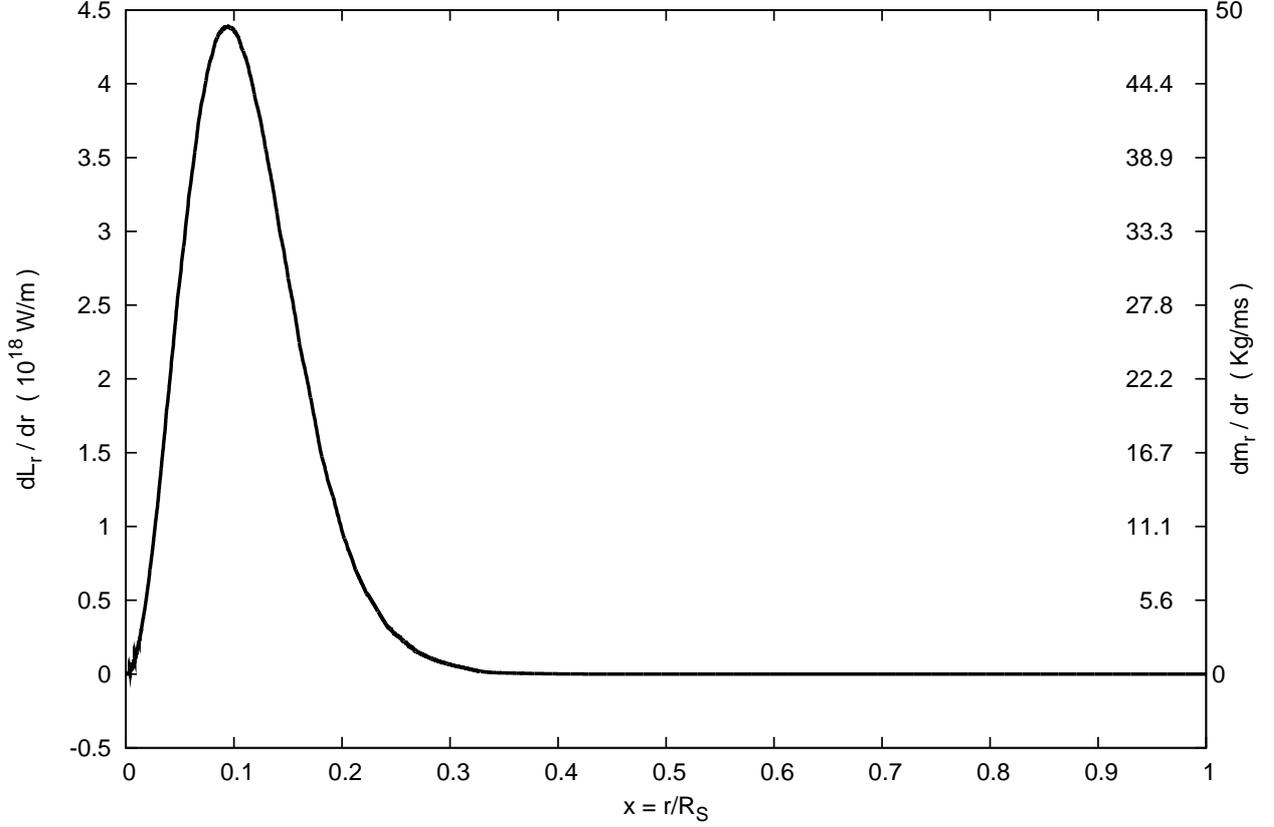}
\par\end{centering}

\caption{Left units: derivative of the interior solar luminosity as a function
of radius (Bahcall et al., 2001; Carroll and Ostlie, 2007). Right
units: derivative of the interior solar burned mass in one second
as a function of radius: see Eq. \ref{eq8000}. A similar shape is
also shown by the amplification conversion function $K(\chi)$, Eq.
\ref{eq819}.}
\end{figure}

\begin{eqnarray}
P_{P}(r/R_{S},t) & = & \frac{3~G~m_{P}~R_{S}^{5}}{2~Q~\Delta t~R_{SP}^{3}}\int_{0}^{r/R_{S}}\chi^{4}\rho(\chi)~\texttt{d}\chi\cdot\label{eq70}\\
 &  & \int_{\theta=0}^{\pi}\int_{\phi=0}^{2\pi}R_{SP}^{3}\left|\frac{\cos^{2}(\alpha_{P,t})-\frac{1}{3}}{R_{SP}^{3}(t)}-\frac{\cos^{2}(\alpha_{P,t-\Delta t})-\frac{1}{3}}{R_{SP}^{3}(t-\Delta t)}\right|~\sin(\theta)~\texttt{d}\theta\texttt{d}\phi,\nonumber
\end{eqnarray}
where $R_{SP}$ is the semi-major axis of a given planet P while $R_{SP}(t)$
is the actual sun-planet distance at the time $t$, the density function
$\rho(\chi)$ is depicted in Figure 6, and $G=6.674\cdot10^{-11}$
N(m/kg)$^{2}$ is the Newton's universal gravitational constant.

Figure 6 also depicts the weighted solar density function

\begin{equation}
\Gamma(r/R_{S})=\int_{0}^{r/R_{S}}\chi^{4}\rho(\chi)~\texttt{d}\chi~.\label{eq55}
\end{equation}
Three values may be of special interests: at the solar core radius,
$\Gamma(0.3)=11.23$ $kg/m^{3}$, inside which solar fusion takes
place; at the solar radiative zone radius, $\Gamma(0.7)=43.96$ $kg/m^{3}$;
and at the solar near surface radius, $\Gamma(1)=51.7$ $kg/m^{3}$.
The core and radiative zone radii are taken from Carroll and Ostlie
(2007).

Note that a comparison between Eqs. \ref{eq32} and \ref{eq70} indicates
that, while the tidal elongation is proportional to the 4th power
of the solar radius, the power associated to the tidal movement is
proportional to the 5th power of the solar radius. This property greatly
increases the importance of tidal work versus tidal elongation. However, the released tidal gravitational power is at least one million times smaller than the TSI fluctuations, which indicates that Newtonian physics alone would rule out any observable planetary influence on the Sun.

\subsection{Nuclear fusion feedback amplification and conversion of tidal gravitational
released  power into TSI at 1 AU.}

Herein we propose a theory based on modern physics. In a given time interval $\Delta t$, nuclear fusion transforms a
certain amount of mass into luminosity that is radiated away. With
the vanished mass, also the gravitational energy of the star would
change by a certain amount $\Delta U_{f}$. However, as each ionized
atom of helium forms from four ionized atoms of hydrogen, the helium
sinks toward the center and space in the core is freed; the pressure
would decrease and the Sun would cool if the freed space is not promptly
filled, and the lost mass replaced by additional sinking hydrogen.
This mass movement releases also a certain amount of potential energy,
$\Delta U_{m}$, as thermal energy to the core. It is possible that
to keep the core fusion activity sufficiently steady, as it is observed,
this complex dynamics occurs in such a way to restore the previous
energetic solar configuration: that is, we can postulate that $\Delta U_{f}+\Delta U_{m}\approx0$.
Thus, there should exist a relation between solar luminosity and the
rate of gravitational power dissipated in the star's core, which, in
turn, should be closely related to the gravitational power associated
to the mass vanished by nuclear fusion activity.

Note that an analogous mechanism is the Kelvin-Helmholtz's contraction
mechanism that continuously causes a warming in the Jovian planets
and in brown dwarfs. For example, Jupiter and Saturn emit almost twice
as much energy as they receive from the Sun. The additional energy
is believed to come from a gradual gravitational contraction and from
the helium slow sinking that continuously convert gravitational potential
energy into thermal energy \citep{Carroll,Bennett}. In the case of
a planet the ratio between the gravitational released power and the
emitted luminosity related to it would be 1, if no amplification from
nuclear fusion occurs. If nuclear fusion occurs, an appropriate amplification
factor $A$ would suffice to describe the phenomenon.

Planetary tides would do two things to the Sun: 1) they would add
a little bit of dissipated power to that already released by the solar
gravity itself, a fact that should slightly increase the average fusion
rate in the core; 2) they add a small oscillation to the gravitational
energy dissipated in the Sun because the tidal amplitude varies in
time, which also forces the nuclear fusion rate to oscillate as well.
An oscillating fusion rate would also force a small expansion/contraction
oscillation of the core, which produces gravitation waves that propagate
fast in the Sun from the core to the convective zone and promptly
transport the energy signal, as it is better explained in the final
paragraph of Section 5.

Point \#1 above suggests the strategy that we need to adopt to solve
the problem. In fact, the tidal work on the Sun would have essentially
the same luminosity increase effect that a star like the Sun, without
planets, would have if its mass is increased by an appropriate small
amount $\Delta M$. The Hertzsprung-Russell diagram for main sequence
stars establishes that, if the mass of a star increases, its luminosity,
$L$, increases as well according to the well-known \emph{mass-luminosity
relation}. In the case of a main-sequence star with mass $M=M_{S}+\Delta M$
close to our solar mass, $M_{S}$, the mass-luminosity relation is
approximately given by

\begin{equation}
\frac{L}{L_{S}}\approx\left(\frac{M}{M_{S}}\right)^{4}\approx1+\frac{4\Delta M}{M_{S}},\label{mlr}
\end{equation}
where $L_{S}$ is the luminosity of our Sun \citep{Duric}. Eq. \ref{mlr}
is good for stellar masses within the interval $0.43<M/M_{S}<2$.
Note that an additional small mass would also imply an additional
gravitational released energy proportional to that mass. As explained
above, we can assume that the Sun works as a system that every second
transforms into luminosity the same amount of gravitational energy
associated to the transformed mass. Thus, we can postulate the existence
of an equation similar to Eq. \ref{mlr} and say that if we indicate
with $\dot{U}_{tidal}$ the small potential power released by tidal
work in the Sun and with $\dot{U}_{Sun}$ the average potential
power released by solar gravity alone as a consequence of the nuclear
fusion activity, the Sun works in such a way to also transform $\dot{U}_{tidal}$
into luminosity according the following relation

\begin{equation}
L(t)\approx L_{S}+4L_{S}~\frac{\dot{U}_{tidal}(t)}{\dot{U}_{Sun}}=L_{S}+A\cdot\dot{U}_{tidal}(t),\label{eq352}
\end{equation}
where $A$ is the amplification factor and $L_{tidal}(t)=A\cdot\dot{U}_{tidal}(t)$
represents the small luminosity anomaly induced by the tidal work
on the Sun, which oscillates in time with the orbits of the planets.
To the first order approximation, the above equation can be used to
convert tidal dissipated work into solar luminosity anomaly once that
$\dot{U}_{Sun}$ is calculated. In the following we check our hypothesis
by calculating the solar response to tidal dissipated energy as described
by Eq. \ref{eq352}, and compare our result to the observed TSI fluctuations.

The total solar luminosity is

\begin{equation}
L_{S}=4\pi(1AU)^{2}\times TSI=3.827\cdot10^{26}~W~,\label{}
\end{equation}
where 1 AU = $1.496\cdot10^{11}$ m is the average Sun-Earth distance,
and TSI is the total solar irradiance, TSI = 1360.94 $W/m^{2}$, at
1 AU. This luminosity is produced mostly in the stellar core. The
luminosity density function is depicted in Figure 7 \citep{Carroll}.
The solar luminosity is generated by nuclear fusion. The function
of the interior mass transformed into energy in one second, $\dot{m}(r)$,
is also indicated by the curve depicted in Figure 7, and it is given
by the well-known Einstein's equation ($E=mc^{2}$):

\begin{equation}
\frac{d\dot{m}(r)}{dr}=\frac{1}{c^{2}}~\frac{dL(r)}{dr},\label{eq8000}
\end{equation}
where $c=2.998\cdot10^{8}~m/s$ is the speed of the light. Because
nuclear fusion converts mass into energy that is radiated away, an
equivalent amount of gravitational potential energy disappears as
well and we assume that, at first approximation, this potential energy
would be compensated by additional gravitational work. Here we postulate
that the rate of total solar gravitational released energy that we
need to use is associated to the lost mass from Eq. \ref{eq8000},
and given by

\begin{equation}
\dot{U}_{Sun}=\frac{1}{2}~G\int_{0}^{R_{S}}m_{S}(r)~\frac{d\dot{m}(r)}{dr}~\frac{1}{r}~\texttt{d}r=3.6\cdot10^{20}~W,\label{eq80}
\end{equation}
where the initial factor $1/2$ is due to the virial theorem. Thus,
using Eq. \ref{eq352}, a first order magnitude of the average amplification
factor for converting gravitational potential released power into
solar luminosity is

\begin{equation}
A=\frac{4L_{S}}{\dot{U}_{Sun}}\approx4.25\cdot10^{6}.\label{eq819}
\end{equation}
The above amplification factor should be distributed inside the Sun,
according to the luminosity density function depicted in Figure 7.
Thus, the function for converting gravitational power into TSI at
a distance $D=1$ AU = $1.496\cdot10^{11}$ m is

\begin{equation}
K(\chi)=\frac{A}{4\pi D^{2}}~\frac{R_{S}}{L_{S}}~\frac{dL(r)}{dr}=\frac{A}{4\pi D^{2}}~\frac{1}{L_{S}}~\frac{dL(\chi)}{d\chi}~,\label{eq819}
\end{equation}
where $\chi=r/R_{S}$. Note that $K(\chi)$ has the same shape of
the function depicted in Figure 7, which stresses the importance of
the solar core, and that $\int_{0}^{1}K(\chi)d\chi=A/(4\pi D^{2})\approx1.6\cdot10^{-17}~m^{-2}$.

\begin{figure}[t]
\begin{centering}
\includegraphics[angle=-90,width=1\textwidth]{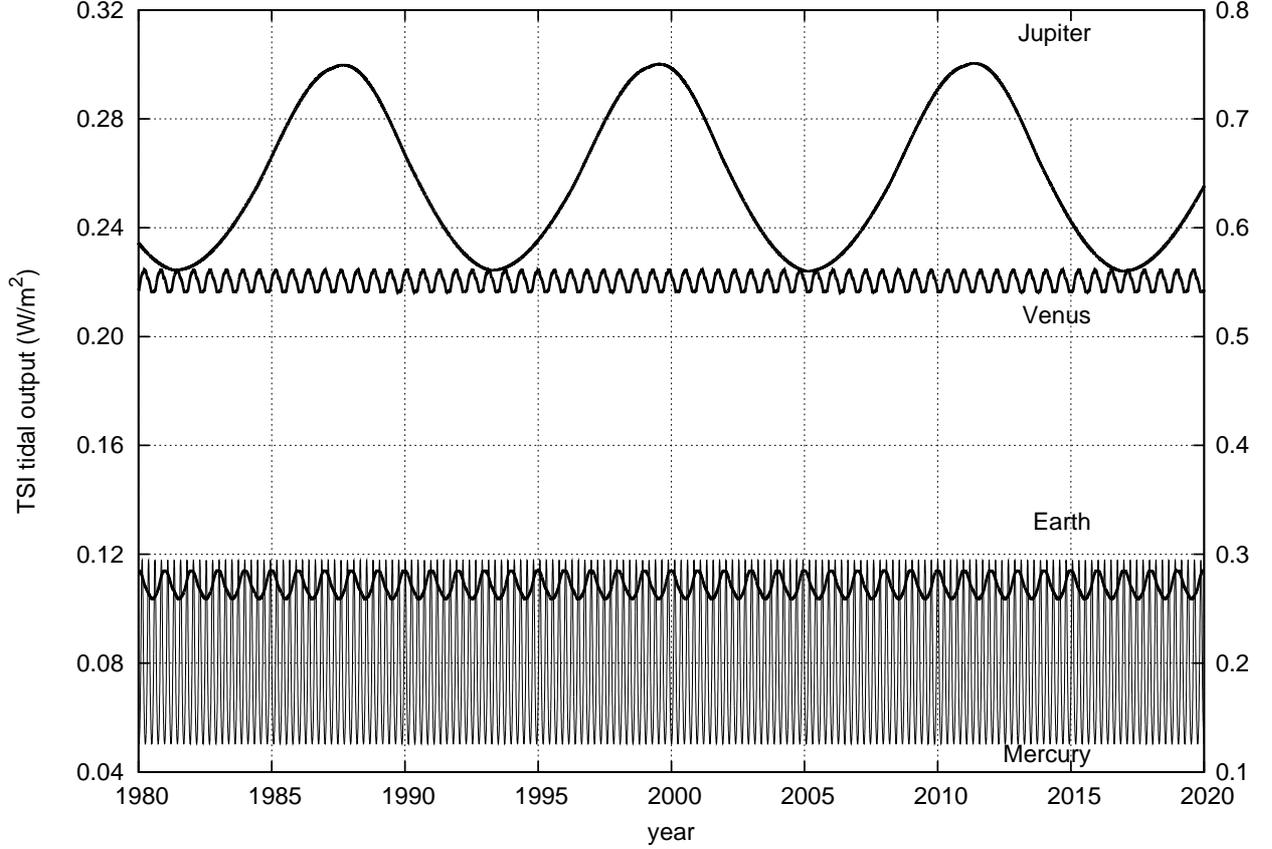}
\par\end{centering}

\caption{Total tidal induced irradiance estimate $I_{P}(1,t)$ according to
Eq \ref{eq91} for Jupiter, Venus, Earth and Mercury. The left scale
refers to the Love number 3/2 and the right scale to 15/4.}
\end{figure}

\begin{figure}[t]
\begin{centering}
\includegraphics[angle=-90,width=1\textwidth]{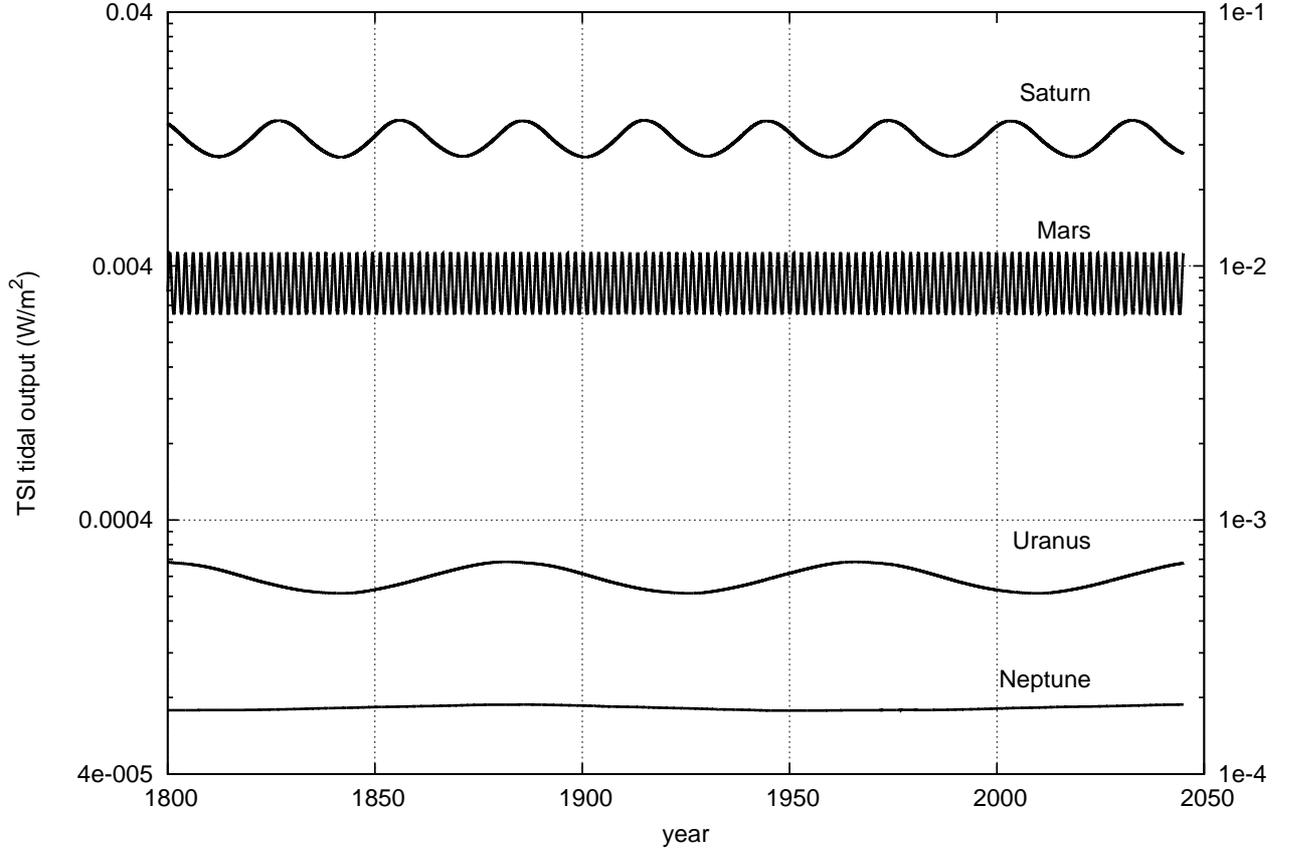}
\par\end{centering}

\caption{Total tidal induced irradiance estimate $I_{P}(1,t)$ according to
Eq \ref{eq91} for Saturn, Mars, Uranus and Neptune.The left scale
refers to the Love number 3/2 and the right scale to 15/4.}
\end{figure}

\subsection{Numerical estimate of the planetary tidal induced TSI oscillations
at 1 AU.}

We combine Eq. \ref{eq70} and Eq. \ref{eq819}, and then integrate
the product function to obtain the total solar irradiance anomaly
function associated to the planetary tide as they evolve in time,
$I_{P}(t)$ . Thus, we obtain

\begin{eqnarray}
I_{P}(t) & = & \frac{3~G~m_{P}~R_{S}^{5}}{2~Q~\Delta t~R_{SP}^{3}}\int_{0}^{1}K(\chi)~\chi^{4}\rho(\chi)~\texttt{d}\chi\cdot\label{eq91}\\
 &  & \int_{\theta=0}^{\pi}\int_{\phi=0}^{2\pi}R_{SP}^{3}\left|\frac{\cos^{2}(\alpha_{P,t})-\frac{1}{3}}{R_{SP}^{3}(t)}-\frac{\cos^{2}(\alpha_{P,t-\Delta t})-\frac{1}{3}}{R_{SP}^{3}(t-\Delta t)}\right|~\sin(\theta)~\texttt{d}\theta\texttt{d}\phi~.\nonumber
\end{eqnarray}
In the following we use the high-precision JPL HORIZONS ephemerides
system to determine the actual daily positions of the planets. The
data are sampled at a daily scale: that is, we choose

\begin{equation}
\Delta t=24*3600=86400~s~.\label{eq92}
\end{equation}
Let us define:

\begin{equation}
I_{P}=\frac{3~G~m_{P}~R_{S}^{5}}{2~Q~\Delta t~R_{SP}^{3}}\int_{0}^{1}K(\chi)~\chi^{4}\rho(\chi)~\texttt{d}\chi=5.2\cdot10^{7}\cdot\frac{m_{p}}{R_{SP}^{3}}\label{eq93}
\end{equation}
The function $I_{P}(t)$ for each planet can be numerically determined
by using Eq. \ref{eq91}. The results are depicted in Figures 8 and
9 and in Table 3. The observed oscillations are due to the elliptical
orbits of the planets with the maximum occurring at the aphelion.
Jupiter produces the largest effect both in average and in the amplitude
of the oscillation. Venus, Earth and Mercury and Saturn follow in
order. The average values $I_{P}$ (see Table 3) vary from about 0.08
$W/m^{2}$ to 0.24 $W/m^{2}$ at least for Mercury, Venus, Earth and
Jupiter. Mars, Uranus and Neptune produce smaller effects. All calculations
need to be amplified by 2.5 if the Love number 15/4 is used, and we
would get a range between 0.2 $W/m^{2}$ and 0.6 $W/m^{2}$ for the
same planets: see Table 3, and Figures 8  and 9. Note that a frequency/amplitude dependency of the factor Q would change the relative weight of the tidal signals but, as explained above, this complex issue will be addressed in future work.

\begin{table}
\begin{centering}
\begin{tabular}{cccccccc}
\hline
Planet  & $I_{P}$  & average  & max-min  & min  & max  & max-date  & period \tabularnewline
 & ($W/m^{2}$)  & ($W/m^{2}$)  & ($W/m^{2}$)  & ($W/m^{2}$)  & ($W/m^{2}$)  & (year)  & (year) \tabularnewline
\hline
Me  & 8.8E-2 (2.2E-1)  & 8.4E-2 (2.1E-1)  & 6.8E-2 (1.7E-1)  & 5.0E-2 (1.3E-1)  & 1.2E-1 (3.0E-1)  & 2000.126  & 0.241 \tabularnewline
\hline
Ve  & 2.0E-1 (5.0E-1)  & 2.2E-1 (5.5E-1)  & 8.8E-3 (2.2E-2)  & 2.2E-1 (5.4E-1)  & 2.3E-1 (5.6E-1)  & 1999.918  & 0.615 \tabularnewline
\hline
Ea  & 9.3E-2 (2.3E-1)  & 1.1E-1 (2.7E-1)  & 1.0E-2 (2.6E-2)  & 1.0E-1 (2.6E-1)  & 1.1E-1 (2.9E-1)  & 2000.011  & 1.00 \tabularnewline
\hline
Ma  & 2.8E-3 (7.1E-3)  & 3.5E-3 (8.9E-3)  & 1.9E-3 (4.7E-3)  & 2.6E-3 (6.5E-3)  & 4.5E-3 (1.1E-2)  & 1999.893  & 1.88 \tabularnewline
\hline
Ju  & 2.1E-1 (5.2E-1)  & 2.6E-1 (6.6E-1)  & 7.6E-2 (1.9E-1)  & 2.2E-1 (5.6E-1)  & 3.0E-1 (7.5E-1)  & 1999.551  & 11.86 \tabularnewline
\hline
Sa  & 1.0E-2 (2.5E-2)  & 1.3E-2 (3.2E-2)  & 4.2E-3 (1.1E-2)  & 1.1E-2 (2.7E-2)  & 1.5E-2 (3.8E-2)  & 2003.258  & 29.46 \tabularnewline
\hline
Ur  & 1.9E-4 (4.8E-4)  & 2.4E-4 (6.0E-4)  & 6.8E-5 (1.7E-4)  & 2.1E-4 (5.2E-4)  & 2.7E-4 (6.9E-4)  & 1965.821  & 84.01 \tabularnewline
\hline
Ne  & 5.8E-5 (1.5E-4)  & 7.3E-5 (1.8E-4)  & 4.0E-6 (1.0E-5)  & 7.1E-5 (1.8E-4)  & 7.5E-5 (1.9E-4)  & 2051.43  & 164.8 \tabularnewline
\hline
\end{tabular}
\par\end{centering}

\caption{Values of $I_{P}$ (Eq. \ref{eq93}), and the average, max-min amplitude,
min, the date of the max closest to 2000 (which is very close to the
aphelion date), and period of the tidal to irradiance estimate function
$I_{P}(t)$ (Eq. \ref{eq91}) for each planet. These curves are reported
in Figures 8 and 9. The values are calculated with the Love numbers
3/2 and 15/4, the latter in parentheses. }

\label{tb3}
\end{table}

\section{Combination of all planetary tides and their $\sim$10-11-12-61 year
TSI induced oscillations}

Herein, we assume that all tides are linearly superimposed and re-write
Eq. \ref{eq91} as

\begin{figure}[t]
\begin{centering}
\includegraphics[angle=-90,width=1\textwidth]{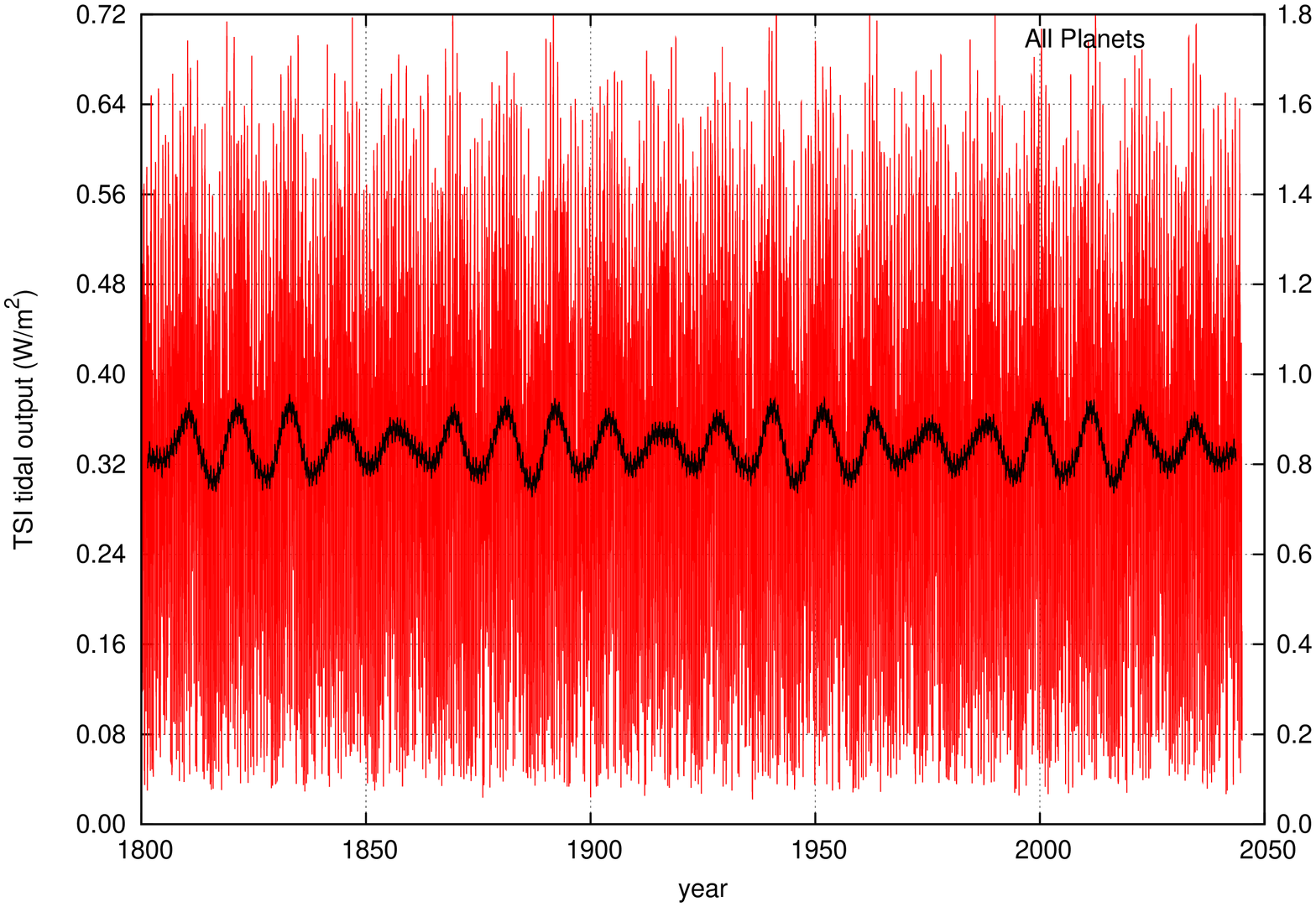}
\par\end{centering}

\caption{Total tidal induced irradiance estimate according to Eq \ref{eq912}
obtained with a linear local superposition of all planetary tides.
A 1001 day moving average curve is added and reveals an approximate
12 year cycle plus a 60-year beat cycle. The left scale refers to
the Love number 3/2 and the right scale to 15/4.}
\end{figure}

\begin{eqnarray}
I_{P}(t) & = & \frac{3~G~R_{S}^{5}}{2~Q~\Delta t}\int_{0}^{1}K(\chi)~\chi^{4}\rho(\chi)~\texttt{d}\chi\cdot\label{eq912}\\
 &  & \int_{\theta=0}^{\pi}\int_{\phi=0}^{2\pi}\left|\sum_{P=1}^{8}~m_{P}\frac{\cos^{2}(\alpha_{P,t})-\frac{1}{3}}{R_{SP}^{3}(t)}-~m_{P}\frac{\cos^{2}(\alpha_{P,t-\Delta t})-\frac{1}{3}}{R_{SP}^{3}(t-\Delta t)}\right|~\sin(\theta)~\texttt{d}\theta\texttt{d}\phi,\nonumber
\end{eqnarray}
where the internal sum is extended to all eight planets. Eq. \ref{eq912}
is numerically integrated, and the output is depicted in Figure 10.
The figure shows very rapid oscillations due to the inner planets.
The average is about $0.35~W/m^{2}$, which is about 250 ppm of the
TSI output. The fast oscillations have a bottom-to-top amplitude up
to about $0.65~W/m^{2}$. Alternatively, if the Love number 15/4 is
used, we would get 0.88 $W/m^{2}$, 1.63 $W/m^{2}$ and 625 ppm, respectively.

A simple 1001-day moving average reveals irregular cycles dominated
by Jupiter's orbital period of 11.86 years with a bottom-to-top amplitude
up to about 0.1 $W/m^{2}$ (or 0.25 $W/m^{2}$, alternatively). In
addition, a beat oscillation of about 61 years is observed. This beat
frequency reveals the presence of another frequency with a period
of about 10 years, which is due to the spring tidal period of Jupiter
and Saturn (see Eq. \ref{beatcc}). The figure shows that these minima
occurred around 1910-1920 and 1970-1980, when solar minima as well
as Earth's global surface temperature minima, are observed \citep{Scafetta20099,Scafetta,scafett2011,Scafetta200}.

\begin{figure}[t]
\begin{centering}
\includegraphics[angle=-90,width=1\textwidth]{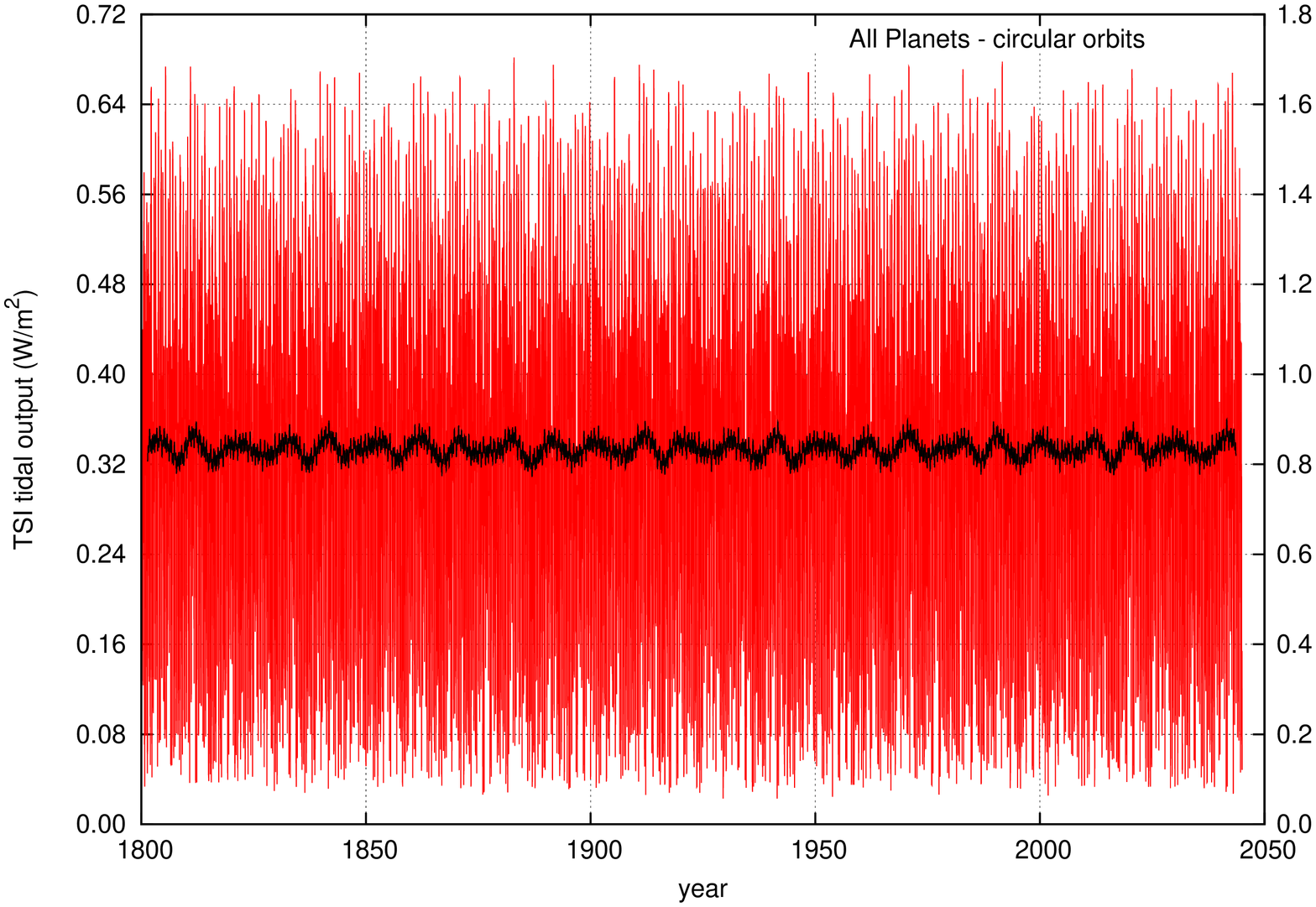}
\par\end{centering}

\caption{Like Figure 10 but now the total tidal induced irradiance is calculated
by keeping the planets at their semi-major axis (sma). A 1001 day
moving average curve is added and reveals an approximate 10 year cycle.
The left scale refers to the Love number 3/2 and the right scale to
15/4.}
\end{figure}

Figure 11 shows an integration of Eq. \ref{eq912}, where the planets
are assumed to have circular orbits at their semi-major axis distance.
The motivation is to study the tidal effect of the planetary orbital
average. In fact, the solar internal dynamics may more easily dampen
the oscillations due to the elliptical shape of the orbits, and, consequently,
may respond more strongly to the tide as calculated at the average
distance of the planets from the Sun, which is given by the semi-major
axis. A simple 1001-day moving average of the integrated record reveals
also irregular cycles, but now the cycles oscillate around a 10-year
period with a max-min amplitude of about 0.05 $W/m^{2}$ (or 0.125
$W/m^{2}$, alternatively).

\begin{figure}
\begin{centering}
\includegraphics[width=0.9\textwidth]{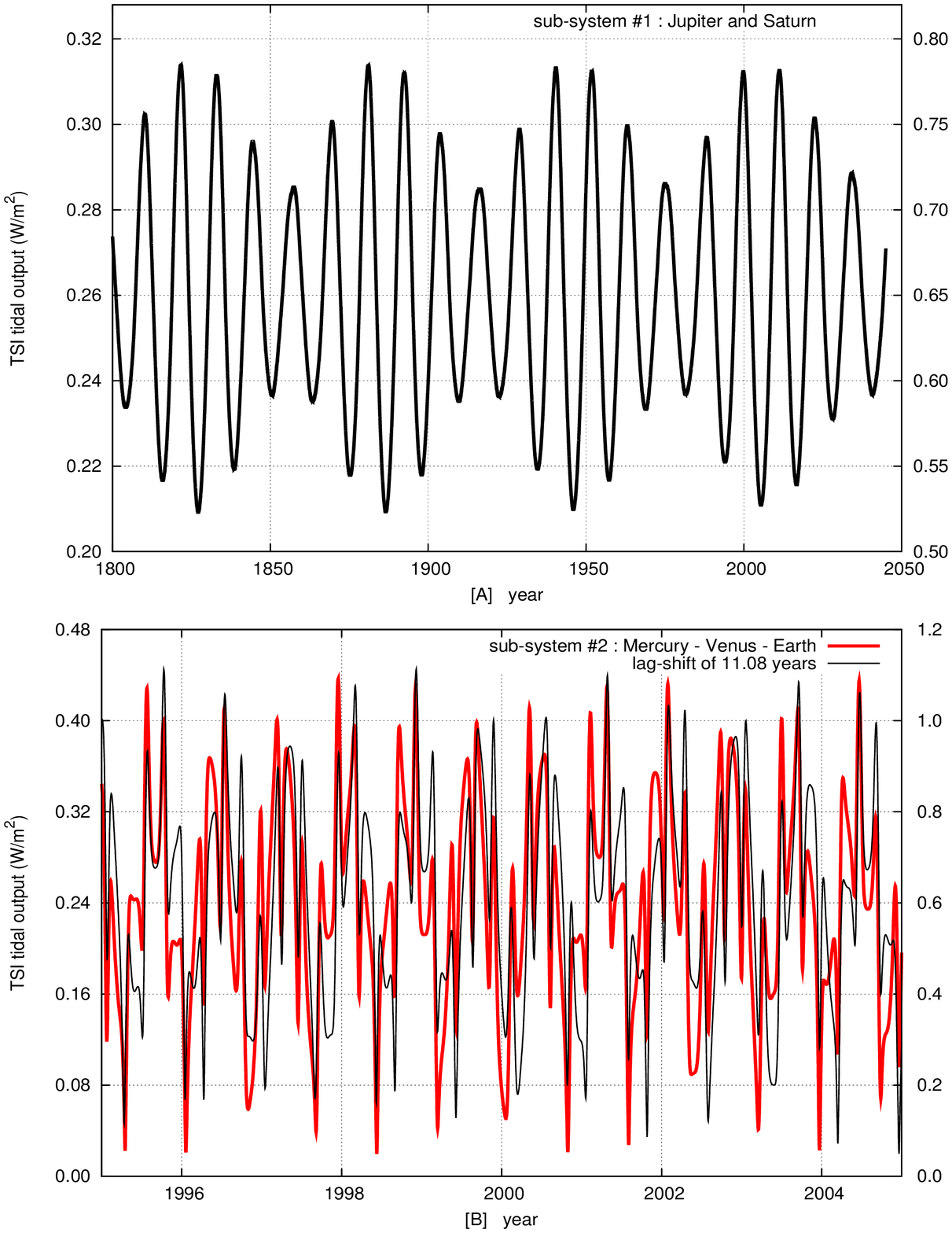}
\par\end{centering}

\caption{Total tidal induced irradiance estimate according to Eq. \ref{eq912}.
{[}A{]} the sub-system Jupiter and Saturn alone: note the $\sim$10-year
(Jupiter-Saturn spring tide) and $\sim$12-year (Jupiter orbit) cycles
that give origin to a $\sim$60-year beat cycle ( see eq. \ref{beatcc}).
{[}B{]} The sub-system Mercury, Venus and Earth; the periods 1995-2005
and 2006.08-2016.08 are superimposed to show an 11.08 year quasi-periodic
recurrence in this tidal pattern. The left scale refers to the Love
number 3/2 and the right scale to 15/4.}
\end{figure}

Finally, Figure 12 depicts the tidal effects of two separated planetary
subsystems: a) Jupiter and Saturn; b) Mercury, Venus and Earth. Figure
12A highlights the existence of the $\sim$10-year Jupiter-Saturn
spring tide cycle and the $\sim$12-year Jupiter orbit cycle that
generate also a $\sim$61-year beat cycle (Eq. \ref{beatcc}). The
first two cycles are clearly seen in the sunspot number record (see
Figure 2), and the $\sim$61-year cycle is found in millennial cosmogenic
isotope records such as $^{14}$C and $^{10}$Be, in the aurora records
\citep{Ogurtsov,Charvatova88,Komitov}, and in numerous terrestrial
climate records \citep{Scafetta,Scafetta200}. Figure 12B highlights
the existence of the 11.08-year recurrent cycle generated by the terrestrial
planets that well agrees with the 11.06-year mean sunspot number cycle
length. The average tidal induced TSI is almost the same for the above
two planetary subsystems: about 0.25 $W/m^{2}$ (or 0.625 $W/m^{2}$,
alternatively). The induced oscillations in the two cases have amplitude:
about 0.1 $W/m^{2}$ and 0.4 $W/m^{2}$ (or 0.25 $W/m^{2}$ and 1.0
$W/m^{2}$, alternatively).

The observed oscillations in Figures 10 and 12 are compatible with
the observed TSI oscillations ($\lessapprox1~W/m^{2}$). Thus, the
tidal induced fluctuations can be considered sufficiently energetic
to activate synchronization and resonance processes yielding a further
adjustment of the signal and/or the emergence of an additional solar
cycle dynamo around 11 years \citep{Scafetta200}. However, as also explained above about the factor Q (Eq. 18), fast and/or large
fluctuations are expected to be attenuated because of damping mechanisms,
which are frequency dependent. These dynamic corrections will be addressed in future work..

Finally, in Figure 13 we compare the Lomb-periodogram spectral analysis
of the sunspot number record and of the total tidal function record
depicted in Figure 10. The figure focus only on the frequency range
of the Schwabe cycle. The Lomb-periodogram of the sunspot record confirms
the finding of the MEM depicted in Figure 2 by showing three peaks
at about 10, 11 and 11.86 years. As expected, the two side frequencies
correspond to the frequencies of  Jupiter/Saturn spring tide (at
about 10 year period) and of  Jupiter orbital tide (at about
11.86). Instead, the central frequency is likely generated by the
solar dynamo which would tend to synchronize to the average beat tidal
frequency as well as at the dynamical recurrence frequency (at 11.08
years) of the fast tides as shown in Figure 12 A. \citet{Scafetta200}
showed that the three cycles beat producing known solar variability
at multiple scales. Again, the frequency/amplitude dependency of the factor Q and of other internal synchronization
and resonance mechanisms may explain why in the observation the $\sim$11.86-year
Jupiter cycle is damped relative to the 9.93-year Jupiter/Saturn spring
tidal signal. Herein, we do not deal with these internal dynamical
resonance mechanisms because we are only interested in determining
whether the energy argument is sufficiently valid to further develop
the theory in the future.

\begin{figure}[t]
\begin{centering}
\includegraphics[angle=-90,width=1\textwidth]{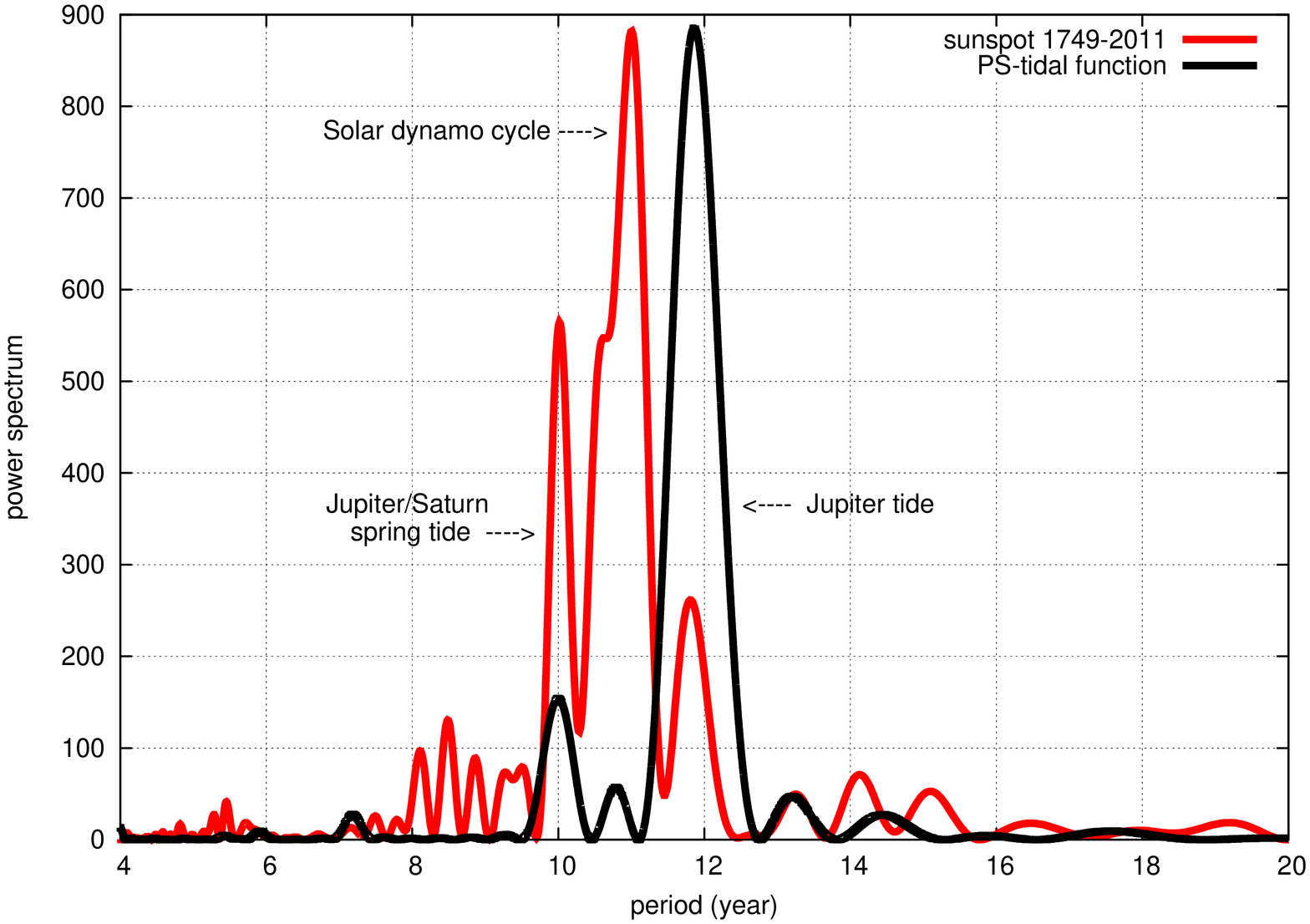}
\par\end{centering}

\caption{ Lomb-periodogram spectral analysis of the sunspot number record and
of the total tidal function record depicted in Figure 10. Note that
the two side frequencies at about 10-year (J/S-spring tide) and 11.86-year
(J-tide) correspond perfectly in the two spectral curves. The central
frequency at about 10.9-year present in the sunspot number spectrum
is generated by the solar dynamo itself while synchronizing its dynamics
to the planetary frequencies. Note that the discrepancy in the relative
amplitude of the side tidal peaks may be due to an internal physical
mechanism that dampens one frequency relative to the other.}
\end{figure}

\section{Rebutting possible objections}

As explained in the Introduction, at least three major objections
exist against a planetary influence on the Sun, and we believe that
they can be rebutted.

The first objection \citep{Smythe} claims that Jupiter and Saturn planetary
tidal patterns would be inconsistent with the periods of prolonged
solar minima such as the Maunder grand solar minimum. This objection
is explicitly addressed and rebutted in \citet{Scafetta200}. There
it is shown how the three frequencies detected in Figure 2 and the
tidal planetary timings can be used to develop an harmonic model capable
of reconstructing all known major solar patterns. These include the
correct timing of the 11-year Schwabe solar cycle and of the known
prolonged periods of low solar activity during the last millennium
such as the Oort, Wolf, Spörer, Maunder and Dalton minima, plus the
emergence and the timing of the observed quasi-millennial solar cycle.

The second  objection is based on Newtonian classical physics alone and claims that the planetary tides are too
small to induce observable effects \citep{Jager,Callebaut}. This paper rebuts
this objection showing that there is the need to take into account modern physics and consider a nuclear fusion feedback
response to tidal forcing that we have calculated to be able to amplify
the gravitational tidal heating effect alone by a million time factor: see Eq. (31).
Moreover, the tides act everywhere inside the Sun, not just at one
level such as at the surface and/or at the tachocline. This implies
that an integration of the tidal effect from the center to the surface
is needed. Consequently, the total tidal effect on the Sun is proportional
to $R_{S}^{5}$ as Eq. \ref{eq91} shows, and not to $R_{S}^{4}$
as normally miscalculated by the critics who calculate the tidal elongation
only at a predetermined distance from the solar center by simply using
Eq. \ref{eq32}. Essentially, the Sun, by means of its nuclear active
core, should work as a great amplifier that increases the strength
of the small gravitational perturbation of the planetary tidal signals
passing through it. We have found that planetary tides may induce
an oscillating luminosity increase from 0.05-0.65 $W/m^{2}$ to 0.25-1.63
$W/m^{2}$. Even in the presence of frequency-dependent damping mechanisms,
which are herein ignored, the estimated luminosity anomaly should
be sufficiently energetic to synchronize solar dynamics with the planetary
frequencies and activate internal dynamical resonance mechanisms,
which then generate and interfere with the solar dynamo cycle to produce
the observed solar variability, as further explained in \cite{Scafetta200}.

The third objection is based on the Kelvin-Helmholtz time scale \citep{Mitalas,Stix}
that would predict that the travel time scale of an erratic photon
from the core to the convective zone ranges between $10^{4}$ to $10^{8}$
years. This argument is used to claim that even if the solar core
gets warmer because of a tidal massaging, the perturbation would reach
the surface on average after hundred thousand years. This time scale
is very long compared to the historical astronomical record, and relatively
small core luminosity variations would be practically smoothed out
and disappear during the very long erratic photon transport journey
to the surface. This topic is not directly addressed in the present
paper because this paper focuses on the tidal heating effect in the
solar core, not on how the energy may be transported to the surface.

Preliminary attempts to solve the above problem have been already
proposed in the scientific literature, where it was assumed that the
solar core is not in a perfect hydrostatic equilibrium because of the tidal heating.
For example, Grandpierre (1990,1996) proposed that planetary tides
induce finite amplitude flows in the core that induce an electric
field generation, which then produces some kind of gently local thermonuclear
runaways which shoot up convective cells to the outer layers. Thermonuclear
runaways processes move energy very fast at a speed of several kilometers
per second, and are well known to cause supernova explosions. More
recently, Wolff and Patrone (2010) argued that: \textit{``an event
deep in the Sun that affects the nuclear burning rate will change
the amount of energy going into the g-mode oscillations. Some information
of this is transported rather promptly by g-modes to the base of the
Sun's convective envelope (CE). Once these waves deposit energy there,
it is carried to the surface in a few months by extra convection,
which should increase solar activity in the way described early in
Section 1. This upward transport of luminosity by waves was also advocated
by Wolff and Mayr (2004) to explain the eastwest reversing flows detected
by Howe et al. (2000) and Komm et al. (2003) with characteristic time
scales of one to three years.''}

Indeed, if the solar nuclear fusion  rate oscillates because of an oscillating
planetary tidal forcing, it should cause gravitational perturbations
through buoyancy waves that should be felt by the entire Sun quite
fast. Thus, it is possible that the energy output variation due to
tidal forcing would propagate through the solar interior by means
of pressure waves at a very high speed. An imperfect analogy would
be given by sound waves and by the pressure propagation in a fluid
which is regulated by Pascal Principle that states that the pressure
applied to an enclosed fluid is transmitted undiminished to every
part of the fluid and that the perturbation propagates with the speed
of the sound in that specific fluid. If the core warms a little bit,
it expands, and all solar interior should rapidly feel the associated
gravitational/pressure effects. These pressure perturbations together
with the increased core luminosity may have the effect of modulating
the luminosity flux toward the tachocline and induce a harmonic modulated forcing of the convective zone that produces the final TSI output. Consequently, the luminosity
output could respond fast to oscillating changes in fusion rate occurring
in the core without the need to wait hundred thousand years so that
the surplus photons produced in the core come out of the radiative
zone.

It is through these wave pressure perturbations that the energy signal
may be transferred from the core to the tachocline quite fast. For
example, the internal solar g-wave oscillations could evolve as: $G(t)=(1+a\cos(\omega_{p}~t))\cos\{[1+b\cos(\omega_{p}~t)]\omega_{g}~t\}$,
where $\omega_{g}$ is the average frequency of the g-waves, $\omega_{p}$
is the planetary induced harmonic modulation, and $a$ and $b$ are
two small parameters. Once at the tachocline, this energy anomaly
would act as a modulating forcing of the solar dynamo and would tend
to synchronize it \citep{Gonzalez,Pikovsky,Scafetta} to its own frequencies
generating a complex Schwabe cycle among other oscillations, as discussed above and in \citet{Scafetta200}.

Finally, there are conflicting studies investigating whether stellar activity
could be strongly enhanced by closely orbiting giant planets \citep{Scharf,Poppenhaeger}. However, until a sufficiently complete planetary-star interaction theory is developed,
the issue cannot be conclusively resolved by simply investigating
other solar systems. In fact, we do not have long enough records nor detailed information about other stars and their solar systems. Moreover, poorly-understood interaction mechanisms,
non-linear effects and stellar inertia mechanisms to fast and large
tidal deformations may lead misleading conclusions. Evidently, a star-planet
interaction theory can only be developed and tested by studying our
Sun and our solar system first, as done herein and in \citet{Scafetta200}.

\section{Conclusion}

Numerous empirical evidences indicate that planetary tides can influence
solar dynamics. High resolution power spectrum analysis reveals that
the sunspot number record presents three frequencies at about Jupiter/Saturn's
spring tidal period of 9.93 years, at 10.87$\pm0.1$ and at Jupiter
period 11.86 years. In addition, the alignment patterns of the sub-systems
of Venus-Earth-Jupiter and Mercury-Venus produce major resonance cycles
at about 11.05-11.10 years, which coincides with the average length
of the observed Schwabe sunspot cycles since 1750. Thus, the Schwabe
solar cycle is reasonably compatible with the tidal cycles produced
by the five major tidal planets: Mercury, Venus, Earth, Jupiter and
Saturn. More details are found in \citet{Scafetta200}, where it is
shown how to reconstruct solar dynamics at multiple time scales using
some of these frequencies.

Despite numerous empirical results, a planetary-solar link theory
has been found problematic in the past mostly because the gravitational
tides induced by the planets on the Sun are tiny, as deduced from
the tidal equation \ref{eq32}. The major tidal planets (Mercury,
Venus, Earth and Jupiter) would produce tides of the order of a millimeter
due to the fact that the tidal elongation is proportional to $R_{S}^{4}$:
see Eq. \ref{eq32}. However, it is the tidal work on the Sun that
physically matters, and we have shown that the total work that the
planetary tides may release to the Sun is proportional to $R_{S}^{5}$.
Indeed, the tides should move up and down the entire column of solar
mass. The tidal movement consistently and continuously squeezes and
stretches the entire Sun from the center to the surface. The solar
mass can be moved and mixed by gravitational tidal forces also because
of the fluid nature of the solar plasma. However, even in this case
only a tiny fraction of the gravitational tidal energy can be released
as heat to the Sun (see Eq. \ref{Q41}), and nothing would be expected
to happen if only released tidal gravitational energy is involved
in the process as Newtonian classical physics  would predict.

However, a planetary tidal \emph{massaging} of the solar core should
continuously release additional heat to it and also favor plasma fuel
mixing. Consequently, the Sun's nuclear fusion rate should be slightly
increased by tidal work and should oscillate with the tidal oscillations.
In section 3.3 we have proposed a methodology to evaluate a nuclear
amplification function (Eq. \ref{eq819}) to convert the gravitational
potential power released in the core by tidal work into solar luminosity.
The strategy is based on the fact that nuclear fusion inside a solar
core is kept active by gravitational forces that continuously compress
the core and very slowly release additional gravitational energy to it, as the
hydrogen fuses into helium. Without gravitational work, no fusion
activity would occur either because the two phenomena are strongly
coupled \citep{Carroll}. Thus, a simple conversion factor should
exist between released tidal gravitational  power and its
induced solar luminosity anomaly.  We have estimated it using a
simple adaptation of the well-known mass-luminosity relation for main-sequence
stars similar to the Sun: see Eq. \ref{eq352}. The average estimated
amplification factor is $A\approx4.25\cdot10^{6}$, but it may vary
within one order of magnitude. In fact, there is uncertainty about
the Love number that in the case of the Sun may be larger than the
used factor 3/2 (see Eq. \ref{eq32}), and the effective tidal dissipation
factor $Q$ likely varies with the tidal frequency and amplitude, and may  be different from the used binary-star average value $Q=10^{6}$ (see Eq. \ref{Q41}).

With the theoretical methodology based on modern physics proposed in section 3.3 we have found
that planetary tides can theoretically induce luminosity oscillations
that are within one order of magnitude compatible with the TSI records. We have
found that planetary tides may induce an oscillating luminosity increase
from 0.05-0.65 $W/m^{2}$ to 0.25-1.63 $W/m^{2}$. Although damping
effects are herein ignored, additional synchronization and resonance
processes may be activated, produce an additional dynamical amplification
effect and the solar dynamo cycle would also contribute to the final
solar cycle as explained in \citet{Scafetta200}. Although these internal dynamic processes are not addressed in this work,    planetary tides appear to be able
to influence solar activity in a measurable way and our results are
consistent with the observations.

Finally, we have shown that the planetary tides produce major cycles
with about 10, 11, 12 and 60 year periods, which correspond to the
cycles observed in the sunspot number record and other solar and climate
records \citep{Ogurtsov,Charvatova88,Komitov, Scafetta}. The cycles
with periods of 10, 12 and 60 years are directly related to Jupiter
and Saturn orbits; the 11-year cycle is the average between the 10-12
year Jupiter-Saturn cycles, and it is also approximately reproduced
by the recurrent tidal patterns generated by the fast tidal cycles
related to Mercury, Venus and Earth. The tidal heating generated by
the two planetary subsystems (terrestrial and jovian planets) is almost
the same. So, terrestrial and jovian planets should be both important
to determine solar dynamics at multiple time scales. In particular
we note from Figure 12A that the combined tides of Jupiter and Saturn
would imply an increased solar activity occurring from 1970 to 2000
with a peak around 2000 that would also be almost in phase with the
10.87-year solar dynamo cycle \citep{Scafetta200}: this pattern would
be qualitatively consistent with the pattern shown by the the ACRIM
total solar irradiance composite depicted in Figure 1 \citep{Willson}.

The preliminary results of this paper suggest that for better understanding
solar activity, the physical interaction between the planets and the
Sun cannot be dismissed, as done until now. Future research should better address the nature of these couplings, which could also be used to better forecast solar activity and climate change
\citep{Scafetta,Scafetta200}. In fact, planetary dynamics can be
rigorously predicted.

\section*{Acknowlegment:}

The author thanks the ACRIMSAT/ACRIM3 Science Team for support.

\end{document}